\documentclass{kapproc} 
\usepackage{graphicx,longtable}

\let\footnote\savefootnote

\let\footnotetext\savefootnotetext 

\let\footnoterule\savefootnoterule 

\kluwerbib

\newcommand{\la}{\mathrel{\vcenter
     {\offinterlineskip \hbox{$<$}\hbox{$\sim$}}}}

\newcommand\apj{ApJ}
\newcommand\apjl{ApJL}
\newcommand\apjs{ApJ Supp.}

\newcommand\aap{A\&A}

\begin{document}

\articletitle{Explosion Mechanisms of Massive Stars}

\articlesubtitle{A Critical Review of Possibilities and Perspectives}

\author{H.-Thomas Janka,\\ 
Robert Buras, Konstantinos Kifonidis, and Markus Rampp}
\affil{Max-Planck-Institut f\"ur Astrophysik,\\
Karl-Schwarzschild-Str.~1, D-85741 Garching, Germany}
\email{thj@mpa-garching.mpg.de}

\author{Tomek Plewa}
\affil{Dept. of Astron. and Astrophysics and
Center for Astrophysical Thermonuclear Flashes,\\
The University of Chicago, Chicago, IL 60637, U.S.A.\\
and
Nicolaus Copernicus Astronomical Center, Bartycka 18, 
00716 Warsaw, Poland}

\begin{abstract}
One of the central problems in supernova theory is the
question how massive stars explode. Understanding the
physical processes that drive the explosion is crucial
for linking the stellar progenitors to the final remnants
and for predicting observable properties like explosion
energies, neutron star and black hole masses, nucleosynthetic
yields, explosion anisotropies, and pulsar kicks.
In this article we
review different suggestions for the explosion mechanism
and discuss the constraints that can or cannot be deduced
from observations. The prompt
hydrodynamical bounce-shock mechanism has turned out not
to work for typical stellar iron cores and empirical values
of the compressibility of bulk nuclear matter.
Magnetohydrodynamical models on the other hand 
contain a number of imponderabilities
and are still far behind the level of refinement that has been
achieved in nonmagnetic simulations. In view of these facts
the neutrino-driven mechanism must still be considered
as the standard paradigm to explain the explosion of
ordinary supernovae, although its viability has yet to be
demonstrated convincingly. Since
spherically symmetric models do not yield explosions, the
hope rests on the helpful effects of convection inside the
nascent neutron star, which could boost the neutrino
luminosity, and convective overturn in the neutrino-heated
region behind the stalled shock, which increases the efficiency
of neutrino-energy transfer in this layer. 
Here we present the first two-dimensional simulations
of these processes which have been performed with a Boltzmann
solver for the neutrino transport and a state-of-the-art
description of neutrino-matter interactions. 
Although our most complete
models fail to explode, convection brings them encouragingly
close to a success. An explosion could be obtained by just a 
minor modification of the neutrino transport, in which
case the exploding model fulfills important requirements
from observations. We discuss necessary improvements on
the route to finally successful models.
\end{abstract}


\section{Introduction}
\label{sec:introduction}

Supernova explosions of massive stars are powered by the
gravitational binding energy that is released when the
initial stellar core collapses to a compact remnant and
its radius shrinks from typically a few thousand kilometers
to little more than ten kilometers.
For solar-metallicity progenitors with main-sequence masses 
of less than about 20--25$\,$M$_{\odot}$ the compact 
leftover will be neutron star. In case of more massive stars
a black hole will be formed, most likely by the fallback 
--- on a timescale of seconds to hours --- of matter that
does not become unbound in the stellar explosion. But also the
direct collapse of the stellar core to a black hole on a 
multiple of the dynamical timescale is possible
(\cite{wooheg02,hegetal02}).

Since the collapse proceeds essentially adiabatically
the total energy of the stellar core is conserved during
the implosion. The gravitational energy is temporarily 
stored as internal energy, mainly of degenerate electrons
and electron neutrinos.  
If rotation plays a significant role in the 
progenitor core, a major fraction of the potential energy 
may also be converted to rotational energy of the nascent
neutron star (or black hole).

The disruption of the massive star in a supernova explosion 
now means that some fraction of the
energy in these reservoirs has to be transferred from the
compact central object to the outer stellar layers.
What are the physical mechanisms to mediate this
energy transfer to the ejecta? And on what timescale do
they work?
Proposals in the literature include the hydrodynamical
bounce-shock, neutrinos, or magnetic fields. The former
would initiate the explosion on a dynamical timescale, 
whereas the
latter two can establish the energy transfer only on the 
secular timescales of neutrino diffusion or magnetic field
amplification, respectively.

Unfortunately, observations have so far not been able to yield
direct insight into the processes in the stellar center at
the onset of the explosion. The hope is that a future Galactic
supernova will change this situation by allowing the 
measurements of a large number of neutrinos and possibly of a
gravitational wave signal in great detail.
The few neutrino events discovered in connection with 
Supernova~1987A were a clear signal of stellar core collapse
and neutron star formation, but they were not sufficient to 
reveal the secrets of the explosion.
Up to now we have to exploit the less direct information that is
provided by the supernova light, by the characteristic properties 
of supernovae and diffuse and compact supernova remnants,
and by the nucleosynthesis of heavy elements which takes 
place in the immediate vicinity of the newly formed neutron
star. 

Section~\ref{sec:observations} will discuss the constraints
for the explosion mechanism that are associated with such 
observations. In Sect.~\ref{sec:theory} we shall briefly review
the different suggestions that have been brought
forward to explain the explosions of massive stars and
will critically evaluate our knowledge on grounds of theoretical
considerations. In Sect.~\ref{sec:neutrinoexplosions} we shall 
summarize the status 
of detailed hydrodynamical supernova simulations and their
implications for our understanding of the delayed explosion
mechanism by neutrino-energy deposition behind the supernova
shock. In Sect.~\ref{sec:results} we shall present the first 
results of a new generation of multi-dimensional supernova
models which employ a Boltzmann solver for the neutrino
transport and a state-of-the-art description of neutrino-matter
interactions. Section~\ref{sec:conclusions} will conclude with 
an appraisal of the results and an outlook on open ends.

\section{Observational Facts}
\label{sec:observations}

Progress in our understanding of the processes that lead
to the explosion of massive stars is mainly based on elaborate
numerical modeling, supplemented by theoretical analysis
and constrained by a growing data base of observed
properties of supernovae. The latter may carry imprints from
the physical conditions very close to the center of the 
explosion. Observable features at very large radii, however, 
can be linked to the actual energy source of the explosion
only indirectly through a variety of intermediate
steps and processes. Any interpretation with respect to
the mechansim that initiates the explosion
therefore requires caution.

A viable model for the explosion mechanism of massive 
stars should ultimately be able to explain the observed 
explosion energies, nucleosynthetic yields (in particular 
of radioactive isotopes like $^{56}$Ni, which are created
near the mass cut), and the masses of the compact remnants
(neutron stars or black holes) and their connection with
the progenitor mass.

\begin{figure}[htb!]
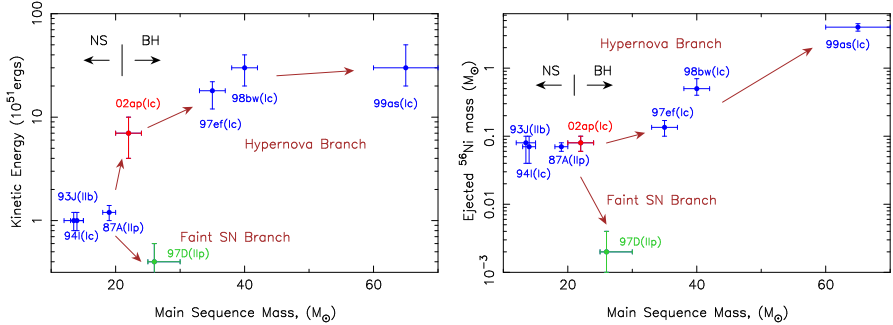

\begin{minipage}[t]{0.48\textwidth}
\begin{center}
\includegraphics[angle=270,width=1.0\textwidth]{Nomoto-1.ps}
\end{center}
\end{minipage}
\hspace{\fill}
\begin{minipage}[t]{0.48\textwidth}
\begin{center}
\includegraphics[angle=270,width=1.0\textwidth]{Nomoto-2.ps}
\end{center}
\end{minipage}
\caption[]{Explosion energy (left) and ejected $^{56}$Ni mass
as functions of the main sequence mass of the progenitor stars
for several supernovae and hypernovae [from \cite{nometal02}].}
\label{fig:observations}
\end{figure}

Recent evaluations of photometric and
spectroscopic data for samples of well-observed Type-II 
plateau supernovae reveal a wide continuum of kinetic energies
and ejected nickel masses. Faint, low-energy cases seem to 
be nickel-poor whereas bright, high-energy explosions tend
to be nickel-rich and associated with more massive 
progenitors (\cite{ham02}). This direct correlation between 
stellar and explosion properties, however, is not 
apparent in an independent analysis by~\cite{nad02}
who speculates that more than one 
stellar parameter (rotation or magnetic fields besides the
progenitor and core mass) might determine the explosion
physics. A large range of nickel masses and explosion
energies was also found for Type Ib/c supernovae (\cite{ham02}).
Interpreting results obtained by the fitting of optical 
lightcurves and spectra, \cite{nometal02} proposed
that explosions of stars with main
sequence masses above 20--25$\,$M$_{\odot}$ split up to a
branch of extraordinarily bright and energetic events
(``hypernovae'') at the one extreme and a branch of faint, 
low-energy or even ``failed'' supernovae at the other. 
Stars with such large masses might collapse to black holes 
rather than neutron stars. The power of the explosion could
depend on the amount of angular momentum in the collapsing
core, which in turn can be sensitive to a number of effects such
as stellar winds and mass loss, metallicity, magnetic fields,
binarity or spiraling-in of a companion star in a binary system.

Another constraint for supernova theory comes from measured
neutron star and black hole masses. A direct mass determination
may not be possible for compact remnants inside supernova
explosions, but the combined information about progenitor
(ejecta) mass, kinetic energy, and nickel mass of a supernova
(as deduced from spectroscopic and photometric data)
may be used to determine the mass cut in a progenitor model by
reproducing the known parameters with numerical simulations.
Masses of binary pulsars (\cite{thocha99,staetal02})
and X-ray binaries (\cite{baietal98})
also add to our knowledge about compact supernova remnants.
In this case, however, binary effects make it even more 
difficult to establish a 
link to progenitor properties and therefore conclusions on the
supernova mechanism should be drawn with great caution.

Complete and incomplete silicon and oxygen burning behind
the supernova shock depend on the strength of the shock
and the structure of the progenitor.
The abundances of corresponding nucleosynthetic products,
if observed in an individual supernova or supernova remnant,
can therefore yield information about the exploding star
and the explosion energy. Their spatial distribution can also
reveal anisotropies and mixing phenomena during the explosion
as in case of, e.g., Supernova~1987A or the
Vela supernova remnant.
Another serious constraint for the explosion mechanism
and the conditions near the mass cut comes
from a comparison of integral yields of supernova
nucleosynthesis with Galactic abundances.
The ejected mass of closed-neutron
shell isotopes with $N=50$ of Sr, Y, and Zr, for example,
is limited to less than about $10^{-4}\,$M$_{\odot}$
per event [\cite{hofwoo96}]. These stable nuclei
are easily created through the capturing of
$\alpha$-particles in the neutron-rich matter around the
mass cut and thus experience conditions in
the close vicinity of the neutron star. The crucial
parameters are entropy and neutron excess. Both
are determined by the exposure to high neutrino fluxes,
because neutrinos exchange energy and lepton number with
the stellar medium.

Anisotropic processes and large-scale mixing between
the deep interior and the hydrogen layer had to be invoked in
case of Supernova~1987A to explain the shape of the lightcurve,
the unexpectedly early appearance of X-ray and $\gamma$-ray
emission, and Doppler features of spectral lines (for a review,
see \cite{nometal94}). Fifteen years after the explosion, the 
expanding debris exhibits an axially symmetric 
deformation \hbox{(\cite{wanwhe02})}.
Supernova~1987A therefore seems to possess
an intrinsic, global asymmetry. The same conclusion was drawn
for other core-collapse supernovae (Type-II as well as Ib/c)
based on the fact that their light is
linearly polarized at a level around 1\% with a tendency
to increase at later phases when greater depths are
observed (\cite{wanhow01,leofil01}).
This has been interpreted as evidence that the inner portions 
of the explosion, and hence the mechanism itself, are strongly
non-spherical (\cite{hoewhe99,wheyi00}),
possibly associated with a ``jet-induced'' 
explosion (\cite{wanwhe02,khohoe99}). This is a very 
interesting and potentially relevant conjecture. 
It does, however, not necessarily
constrain the nature of the physical process that mediates the 
energy transfer from the collapsed core of the star to the
ejecta and thereby creates the asphericity.

\begin{figure}[htb!]
\begin{center}
\includegraphics[width=0.82\textwidth]{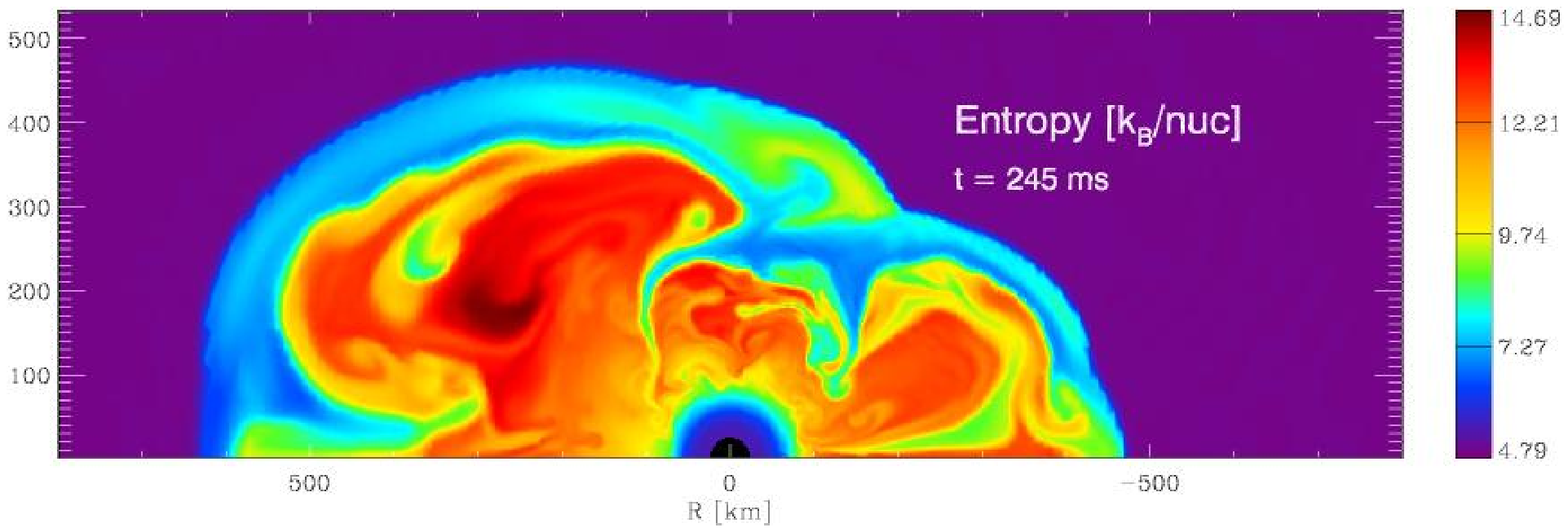}
\end{center}
\vspace{-5mm}
\begin{center}
\includegraphics[width=0.82\textwidth]{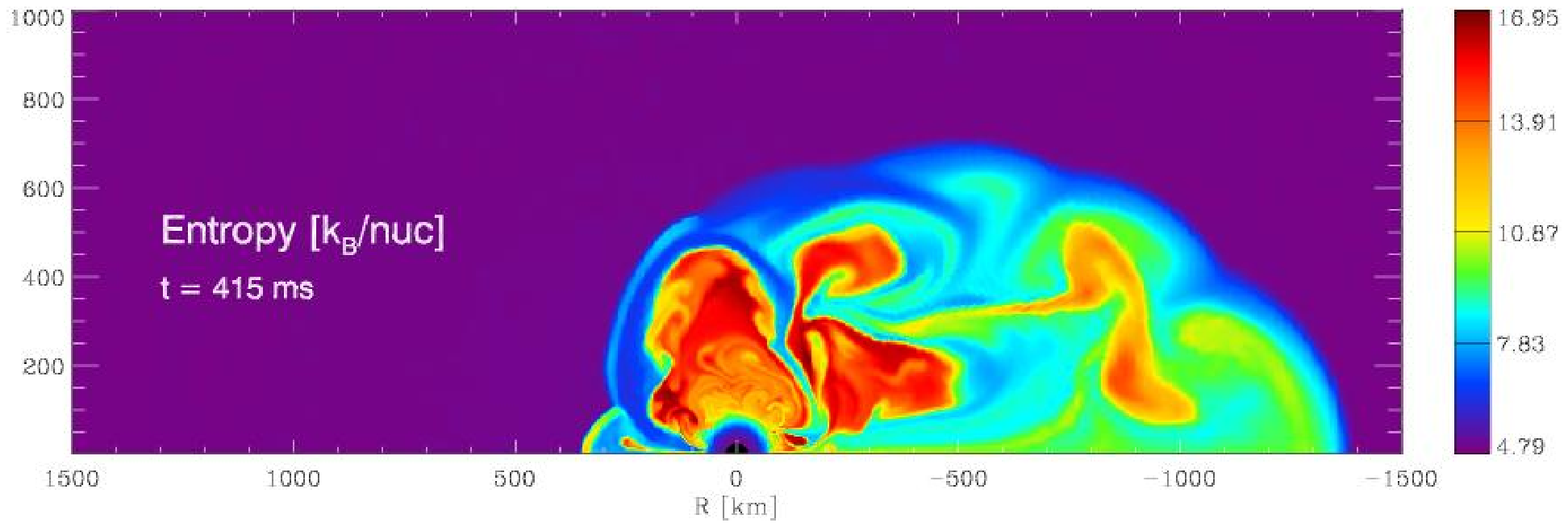}
\end{center}
\vspace{-5mm}
\begin{center}
\includegraphics[width=0.82\textwidth]{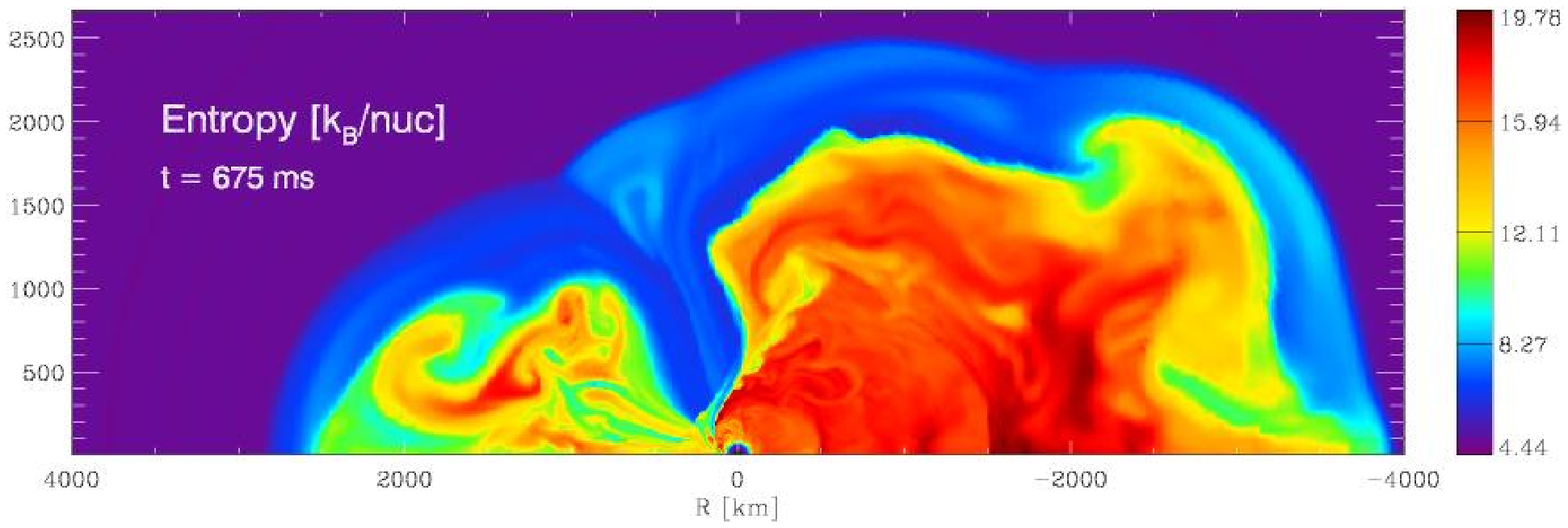}
\end{center}
\vspace{-5mm}
\begin{center}
\includegraphics[width=0.82\textwidth]{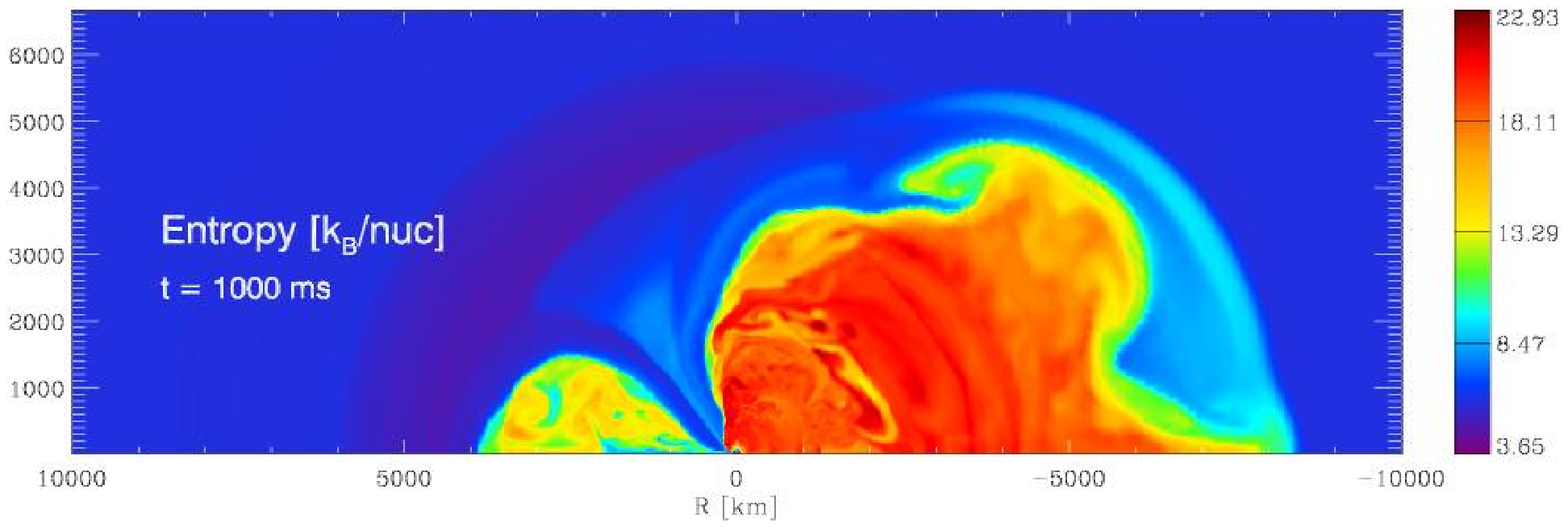}
\end{center}
\caption[]{Explosion that is driven by neutrino-energy deposition 
in combination with convective overturn in the region
behind the supernova shock. The anisotropy of the neutrino-
and shock-heated ejecta is growing in time and becomes very
large due to an increasing contribution of the $m=0$, $l=1$ 
mode in the convective pattern. The snapshots (from top to 
bottom) show the entropy distribution (values between
about 4 and 23$\,k_{\mathrm B}$ per nucleon) at post-bounce times
$t_{\mathrm{pb}} = 245\,$ms, 415$\,$ms, 675$\,$ms,
and 1000$\,$ms. 
Note that the radial scales of the figures differ.
The neutron star is at the origin of
the axially symmetric (2D) grid and plays the role of an
isotropic neutrino ``light bulb'' (Plewa et al., in preparation).} 
\label{fig:monomode}
\end{figure}

The high space velocities of young pulsars with a mean
around 400$\,$km/s (\cite{lynlor94,corche98})
have also been used to argue for an intrinsically asymmetric
explosion mechanism. It is very likely that the neutron
stars obtained their recoil at the onset of the explosion or shortly
afterwards (for a review, see \cite{lai01,laiche01}).
But it is not clear whether the mechanism and the anisotropies of
the explosion are directly connected with the pulsar kicks.
Also a possible ``jet'' and perhaps a ``counter jet'' in
the gaseous filaments of the Cas~A supernova remnant
may neither be linked to the process that
started the supernova explosion nor to the runaway velocity
of the compact remnant. The latter could have been accelerated
(maybe by anisotropic neutrino emission) {\em after} the explosion
was launched. The jet-like features --- if at all caused by processes
at the supernova center and not by local inhomogeneities at the
forward shock --- may have been produced
even later, for example by bipolar outflow from a neutron star
or a black hole that has accreted from a disk of fallback matter.
Interestingly, inspecting the images given by \cite{gotetal01},
the dislocation of the compact source relative
to the geometrical center of the remnant appears not to coincide
with the direction of the possible ``jets'' (compare also
\cite{thoetal01}).

Rotation plus magnetic fields were proposed as the 
``most obvious'' way to break the spherical symmetry and to
explain the global asphericity of core-collapse 
supernovae (\cite{whemei02,akietal02,whe02}).
It was argued that current numerical calculations may be 
missing a major ingredient necessary to yield explosions.
A proper treatment 
of rotation {\em and} magnetic fields may be necessary to 
fully understand when and how stellar collapse leads to an explosion.
Of course, this might be true. But a confirmation or 
rejection will require computer models with ultimately the full
physics. 

It must be stressed, however, that current observations
do not necessitate such conclusions and hydrodynamical
simulations suggest other possible explanations. 
Strong convection in the
neutrino-heating region behind the supernova shock can account
for huge anisotopies of the inner supernova ejecta, even 
without invoking rotation. If the explosion occurs quickly,
much power remains on smaller scales until the expansion sets in
and the convective pattern gets frozen in. If, in contrast, the 
shock radius grows only very slowly and the explosion is delayed
for several 100$\,$ms after bounce, the convective flow can
merge to increasingly larger structures. 
In two-dimensional (2D) hydrodynamic calculations including 
cooling and heating by neutrinos between the neutron star and
the shock (with parameter choices for a central, {\em isotropic} 
neutrino ``light bulb'' which enabled explosions),
we (Plewa et al., in preparation) found situations
where the convective pattern revealed a contribution of the $l=1$,
$m=0$ mode that was growing with time and was even dominant at
about one second after bounce (Fig.~\ref{fig:monomode}). 
\cite{her95} already speculated about such a possibility.
Certainly three-dimensional (3D) calculations of the full sphere
(and without the coordinate singularity on the axis of the
spherical grid) are indispensable to convincingly demonstrate 
the existence of this phenomenon\footnote{Another interesting 
possibility was pointed out by \cite{bloetal02}.
Their 2D (and 3D; \cite{mez02}) calculations
revealed hydrodynamical instabilities in the accretion flow
behind the stalled shock, which instigate large-scale modes 
even in the absence of neutrino heating. Whether these
calculations are of relevance for the supernova problem 
has be to seen.}.

\section{Theoretical Possibilities}
\label{sec:theory}

The primary energy source for powering supernovae of massive
stars is the gravitational binding energy of the newly formed
proto-neutron star or proto-black hole
(energy from nuclear reactions contributes at a minor
level). To initiate and drive the explosion, energy 
has to be transferred to the stellar matter above the mass 
cut to be finally converted to kinetic  
energy of the ejecta. This could be achieved by
hydrodynamical shocks, neutrinos,
or magnetic fields as mediators. Collimated outflows or jets of
relativistic plasma might also play a role for very special 
circumstances. Depending on the mediator, the
conditions for efficient energy transfer, the corresponding
timescale, and the tapped energy reservoir are different.
A lot of work has been spent in the past 40 years to
identify viable supernova mechanisms and to study the
involved physics. The space here is not sufficient to review
all important papers. We shall therefore
only focus on some key contributions without attempting
completeness.

\subsection{Hydrodynamical (Prompt) Mechanism}

When the homologously
collapsing inner part of the stellar core is abruptly
stopped at core bounce by the stiffening of the nuclear
equation of state (EoS), a hydrodynamical shock is formed
at the boundary to the supersonically infalling outer part
(\cite{colwhi66}). In this case the initial
energy of the shock is equal to the kinetic energy of
the inner core ``at the last moment of good homology''
(\cite{moe93}): 
\begin{equation}
E_{\mathrm{sh,i}}\,\approx\, 6\!\times\! 10^{51}\,{M_{\mathrm{ic}}
\over {\mathrm M}_{\odot}}\,\left (v_{\mathrm{max}}\over 4\!\times\! 10^9\,
{\mathrm{cm\,s}}^{-1}\right )^{\! 2}\ \,{\mathrm{erg}}\, ,
\end{equation}
when $M_{\mathrm{ic}}$ is the mass of the homologously 
collapsing part of the stellar core and $v_{\mathrm{max}}$
its maximum infall velocity.
There is general agreement now, supported by a large
number of numerical calculations (e.g., 
\cite{bru85,bru89a,bru89b,myrblu87,myrblu89,sweetal94}),
that this energy is not
sufficient for the shock to reach the surface of the iron
core: The energy losses from behind the shock by 
photodisintegration
of nuclei to $\alpha$-particles and free nucleons
($\sim 1.5\times 10^{51}\,$erg per 0.1$\,$M$_{\odot}$)
as well as neutrino escape after shock breakout
through the neutrinosphere (some $10^{51}\,$erg) are so severe
that the shock stalls within milliseconds.
Only in case of extremely small iron cores
($M \la 1.1\,$M$_{\odot}$;\cite{barcoo90})
or an extraordinarily soft nuclear EoS (\cite{barcoo85,baretal87})
is the prompt hydrodynamical shock able to disrupt the star.
Nevertheless, the hope that the prompt hydrodynamical
bounce-shock mechanism might work for some stars in a mass window
around 10$\,$M$_{\odot}$ has not been given up completely
(\cite{hiletal84,sumetal01}).

\subsection{Magnetohydrodynamical (MHD) Mechanism}

Soon after the discovery of pulsars the role of rotation and
magnetic fields in the supernova explosion was scrutinized
(\cite{ostgun71,biskog71}).
Because of flux conservation a seed field in the stellar
core can grow significantly during collapse, although for
realistic initial fields not to a strength that magnetic
pressure could be dynamically important. But winding of
field lines in the case of differential rotation, which is
natural after the collapse of a spinning core,
can further amplify the toroidal field component. If the
magnetic pressure becomes comparable to the thermal pressure,
magnetohydrodynamical forces can drive an explosion
(\cite{muehil79}), and might accelerate axial
jets (\cite{lebwil70,whemei02}). Also
buoyancy instabilities of highly magnetized matter could
produce mass motions (\cite{meietal76}).
In this case rotational energy of the spinning proto-neutron
star is converted to magnetic energy, a part of which
might then end up as kinetic energy of the ejecta. Hence,
\begin{equation}
E_{\mathrm{ej}}\,\la \,5\!\times\!10^{52}\,\,{M_{\mathrm{ns}}\over 
1.5\,{\mathrm{M}}_{\odot}}
\,\left ({R_{\mathrm{ns}}\over 10\,{\mathrm{km}}}\right )^{\! 2}
\left ({T_{\mathrm{rot}}\over
1\,\mathrm{ms}}\right )^{\! -2}\ \,{\mathrm{erg}}\,, 
\end{equation}
when $T_{\mathrm{rot}}$ is the
spin period of the forming neutron star with 
mass $M_{\mathrm{ns}}$. The field,
however, grows linearly with time and therefore the timescale
to reach sufficiently strong fields is long compared to
the accretion timescale of the collapsed core, unless
unrealistically large initial values are assumed 
(\cite{meietal76,muehil79}). Recently it was
suggested that the magnetorotational instability (MRI), which
was discussed in the context of accretion disks by 
\cite{balhaw98}, might also be in action in the supernova
core (\cite{akietal02}). Because of an exponential growth,
this would reduce the field amplification timescale drastically
and would allow for much smaller initial seed fields, too.
It remains to be shown, however, whether the field can grow
to a dynamically effective strength in a self-consistent
model for reasonable assumptions
about the rotation of stellar cores and despite of diverse
effects that can lead to field saturation or hinder field
growth. Moreover, it is unclear whether axial jets can
be initiated by the MRI at the conditions in collapsing
stellar cores. Implications for supernova nucleosynthesis
are unexplored, but might be problematic
if MHD explosions are frequent (\cite{meietal76}).
The amplification and transport of magnetic fields in
convective regions inside the nascent neutron star 
(\cite{thodun93}) and in the neutrino-heating layer behind the
supernova shock (\cite{thomur01}) also deserve further
attention with respect to the generation of pulsar fields.

\subsection{Neutrino-Heating (Delayed) Mechanism}

While neutrinos drain energy from the shock-heated
matter immediately after bounce, the situation changes
somewhat later when the density and temperature behind the
shock have dropped. High-energy neutrinos, emitted abundantly
from the hot accretion layer that covers the 
nascent neutron star, are absorbed with a small, but finite
probability (typically $\sim 10$\%)
in the postshock gas. If this energy deposition is
large enough, the stalled supernova shock can be revived.
Since this can happen on a timescale of hundred or more
milliseconds after bounce, the explosion is called ``delayed''
in contrast to the prompt bounce-shock mechanism
(\cite{wil85,betwil85}). The idea that the
energy transfer to the ejecta is mediated by
neutrinos goes back to (\cite{colwhi66}).
During core collapse and subsequent
contraction, the gravitational binding energy of the
forming neutron star is first stored as internal energy
(degeneracy energy of leptons and thermal energy of nucleons)
and only slowly carried away by neutrinos on the diffusion
timescale of several seconds. The kinetic energy required for
the supernova explosion is of order $10^{51}\,$erg and hence
tiny compared to the huge reservoir of binding energy,
\begin{equation}
E_{\mathrm{kin}}\, \ll\, {3\over 5}\,{GM_{\mathrm{ns}}^2
\over R_{\mathrm{ns}}}
\,\sim \,{1\over 10}\,M_{\mathrm{ns}}c^2 \,\sim\, 3\!
\times\!10^{53}\,\,{\mathrm{erg}}\,,
\end{equation}
where the second expression results from the first by using
$R_{\mathrm{ns}}\sim 3R_{\mathrm{s}} \equiv 6GM_{\mathrm{ns}}/c^2$.
But just a smaller fraction (about 10\%) of this energy is actually
radiated in neutrinos during the first hundreds of milliseconds 
until the explosion starts. 

Neutrino heating behind the shock is essentially {\em unavoidable}.
The temperature in this region drops roughly like
$T\propto r^{-1}$ (close to isentropic conditions),
which means that the energy loss rate of the stellar plasma
by neutrinos produced in $e^\pm$ captures on
protons and neutrons falls like
$Q_{\nu}^- \propto T^6 \propto r^{-6}$. In contrast, the
energy deposition by $\nu_e$-absorption on neutrons and
$\bar\nu_e$-absorption on protons decreases much more
slowly. The heating rate varies roughly proportional to
the neutrino energy density (which scales with the summed
luminosity of $\nu_e$ plus $\bar\nu_e$, $L_{\nu}$, divided
by the square of the inverse distance $r$ from the
neutrinosphere) and the mean squared energy
$\left\langle\epsilon_{\nu}^2\right\rangle$ (because
of the dependence of the absorption cross section on the
neutrino energy $\epsilon_{\nu}$),
i.e., $Q_{\nu}^+ \propto L_{\nu}\left\langle
\epsilon_{\nu}^2\right\rangle/r^2$. This means that there
must be a radial position, the so-called ``gain radius''
(\cite{betwil85}), exterior of which neutrino heating
dominates neutrino cooling. The heating rate for matter
that is disintegrated to free neutrons and protons can be
estimated as 
\begin{equation}
Q_{\nu}^+\, \sim\, 550\,\, {L_{\nu,53}
\langle \epsilon_{\nu,15}^2\rangle \over r_7^2}\ \,
{\mathrm{{MeV\over s}}}
\end{equation}
per nucleon for a typical luminosity of $\nu_e$ plus $\bar\nu_e$ 
of $10^{53}\,$erg/s, 15$\,$MeV for the rms neutrino energy,
and $r$ measured in units of $10^7\,$cm. The explosion energy
should scale as 
\begin{equation}
E_{\mathrm{exp}}\,\sim\,
{\Delta M\over m_{\mathrm{u}}}\,\,Q_{\nu}^+\Delta t_{\mathrm{h}}
\,\,,
\end{equation}
where $\Delta M$ is the neutrino-heated mass, $m_{\mathrm{u}}$
the nucleon mass, and $\Delta t_{\mathrm h}$ the timescale
of neutrino-energy deposition. The latter can be assumed to be
roughly the time until nucleons have absorbed an energy
comparable to their binding energy in the
gravitational potential of the neutron star, i.e.,
\begin{equation}
\Delta t_{\mathrm{h}}\,\sim\, {GM_{\mathrm{ns}}m_{\rm u}
\over Q_{\nu}^+\, r}\, \sim\, 40\,\ {\mathrm{ms}} 
\end{equation}
for $M_{\mathrm{ns}}\approx 1.5\,$M$_{\odot}$. These arguments
suggest that the gradual (in contrast to impulsive) energy 
deposition by neutrinos might not be responsible for a 
dominant fraction of the net explosion energy of a supernova.
Instead, neutrino heating helps to compensate the gravitational 
binding and thus triggers the outward expansion of the supernova
ejecta. The energy per baryon increases further by ongoing 
energy transfer 
from neutrinos to the accelerating matter, and finally reaches 
its limiting value due to the energy release in recombinations
of free nucleons to nuclei. This recombination energy may account
for a major part of the supernova energy. 
Numerical simulations and analytic considerations (\cite{jan01})
suggest a mass $\Delta M$ between a few $10^{-2}\,$M$_{\odot}$
and about $10^{-1}\,$M$_{\odot}$. With a typical energy deposited
by neutrino heating being around 5$\,$MeV per nucleon 
(\cite{jan01}), the explosion
energy could range between some $10^{50}\,$erg and several
$10^{51}\,$erg, when about 8$\,$MeV are set free for each 
nucleon that recombines. These numbers naturally match the 
energies of observed normal supernovae
(cf.~Sect.~\ref{sec:observations}).

\subsection{Jet-Powered Explosions}

Stars with main sequence masses beyond 20--$25\,$M$_{\odot}$
seem to be associated with much more powerful explosions
with energies up to several $10^{52}\,$erg
(Sect.~\ref{sec:observations}). Such energies are probably
out of reach for the neutrino-heating mechanism as described
above. The idea has been coined (\cite{popetal99,macfwoo99})
that the core of such stars collapses to
a black hole, which then continues to accrete the infalling
matter of the progenitor star (``collapsar'' or
``failed supernova''; \cite{woo93}). Provided that the stellar core
has retained enough angular momentum during its pre-collapse
evolution (or was spun up by a merger with a binary companion),
the accretion will proceed at very high rates through a disk or
torus on the timescale of viscous angular momentum transport,
which is much longer than the dynamical timescale. Therefore 
the efficiency of energy release can in principle be much higher
than in case of spherical accretion. In terms of the 
rest-mass energy of the accreted matter one can get
$E_{\mathrm{acc}} = \xi M_{\mathrm{acc}}c^2$ with
maximum values between $\xi = 0.057$ for a non-rotating black
hole and $\xi = 0.42$ for an extreme Kerr hole that accretes from
a corotating thin disk (\cite{shateu83}). 
Since the mass accretion rates are
hypercritical and the densities in the disk
correspondingly high, photons cannot escape and the energy
is set free in form of neutrinos rather than electromagnetic
radiation (\cite{popetal99}). The efficiency may
be reduced compared to the quoted maximum values
if the density and temperature are too low for rapid neutrino
production or the disk is very dense and thus becomes 
nontransparent to neutrinos. In these cases the energy is advected
into the black hole by the accretion flow instead of being
carried away by neutrinos (\cite{dimetal02}). 

Alternatively,
magnetohydrodynamical processes may drive outflows from the
disk (e.g., \cite{meietal01,daimoc02,dre02,drespr02}).
It is also possible that the rotational energy of the spinning
black hole is extracted via the Blandford and Znajek mechanism
(\cite{blazna77}) through magnetic fields that are
anchored in the black hole and couple it to the surrounding
disk.

By the annihilation of neutrinos and antineutrinos to
$e^\pm$-pairs (\cite{woo93}) or the mentioned MHD processes,
polar jets could be launched to propagate through the star
along the rotation axis (\cite{macfwoo99,aloyetal00,zhaetal02}).
In case the jets remain collimated and
highly relativistic until they break out of the stellar
surface (which requires that the star does not possess an
extended hydrogen envelope; \cite{mat02}) gamma-ray bursts (GRBs)
may be triggered. Nonrelativistic jets that expand laterally 
and sweep up the surrounding progenitor gas, or energetic winds
from the accretion torus might be the driving
force behind enormously powerful stellar explosions
(\cite{macfetal01}) which
are observed as so-called ``hypernovae'' and GRB-supernovae
(\cite{iwaetal98}). For an overview and more information, see 
\cite{wooetal02}.

\bigskip
The different mechanisms for massive star explosions as 
described above
require different properties of the progenitor stars and
different physical conditions in the collapsed stellar cores.
The latter are incompletely known, but some of the 
requirements are more likely fulfilled than others, 
some combinations of necessary conditions may be more
common and more typical, while others might be realized
only in rare cases and for very special, exceptional 
circumstances. It is possible, if not probable, that the 
diversity of the observed explosions of massive stars and the 
large number of associated phenomena mean that more than
one mechanism is at work. It may also be,
however, that the majority of ordinary supernovae
can be ascribed to the same central process which might
then depend extremely sensitively on yet to be
determined parameters.

\section{Do Neutrino-Driven Explosions Work?}
\label{sec:neutrinoexplosions}

The neutrino-driven mechanism (\cite{wil85,betwil85}) involves
a minimum of controversial assumptions and uncertain
degrees of freedom in the physics of collapsing stars.
It relies on the importance of neutrinos and their
energetic dominance in the supernova core. After the detection
of neutrinos in connection with Supernova~1987A and the
overall confirmation of theoretical expectations
for the neutrino emission, this is not a speculation any more
but an
established fact. Of course, this does not mean that
such a minimal input is sufficient to understand the cause
of supernova explosions and to explain all observable
properties of supernovae. But at least it can be taken as a   
good reason to investigate how far one can advance with a
minimum of imponderabilities. Detailed numerical simulations
are needed to answer the crucial question in this context:
Is neutrino-energy deposition efficient enough to drive the
explosion?

Spherically symmetric simulations with the current input
physics (neutrino interactions and the equation of state of
dense matter) do not yield explosions by the neutrino-heating
mechanism. There is no controversy about that. All computations
are in agreement, independent of Newtonian or relativistic
gravity and independent of the neutrino transport being treated
in an approximate way by flux-limited diffusion 
methods (e.g., \cite{myrblu87,myrblu89,bru93,brunis01}) or
very elaborately by solving the frequency- and angle-dependent
Boltzmann transport 
equation (\cite{ramjan00,mezlie01,liemez01,liemes02}).

Whether neutrinos succeed in reviving the stalled shock
depends on the efficiency 
of the energy transfer to the postshock layer, which in turn
increases with the neutrino luminosity and the hardness of
the neutrino spectrum. 
Wilson and collaborators (\cite{wilmay88,wilmay93,maytav93,totsat98})
have obtained explosions in one-dimensional (1D) simulations for more 
than ten years now. In these models it is, however, {\em assumed} that
neutron-finger convection in the hot neutron star boosts the 
neutrino luminosities. Moreover, a special equation of state
with a high abundance of pions in the nuclear matter was used
(\cite{maytav93}), which again leads to higher neutrino
fluxes from the neutron star and thus to enhanced 
energy-deposition behind the shock. Both assumptions
are not generally accepted.

Two-dimensional 
(\cite{herben92,herben94,shiyam94,burhay95,janmue96,mezcal98:ndconv,shiebi01})
and 3D simulations (\cite{shiyam93,frywar02}) as well as
theoretical considerations (\cite{bet90}) have shown
that the neutrino-heating layer is unstable to
convective overturn. The associated effects support shock
revival and can lead to explosions even in cases where 
spherical models fail. In the multi-dimensional situation
accretion and expansion can occur simultaneously.
Downflows of cooler,
low-entropy matter that has fallen through the shock, 
coexist with rising bubbles of high-entropy, neutrino-heated 
gas (Fig.~\ref{fig:monomode}). 
On the one hand, the downflows carry cool material close
to the gain radius where it absorbs energy readily from the
intense neutrino fluxes. On the other hand, heated matter rises
in bubbles and can expand and cool quickly, which reduces
the energy loss by the reemission of neutrinos. This also
increases the postshock pressure and hence pushes
the shock farther out. The gain layer grows and thus more
gas can accumulate in the neutrino-heating region.
The gas also stays longer in the gain layer, in
contrast to one-dimensional models where the matter behind the 
accretion shock has negative velocity and is quickly advected
down to the cooling layer. When the gas arrives there, 
neutrino emission
sets in and extracts again the energy which had been absorbed 
from neutrino heating shortly before. Due to the combination of
all these effects postshock convection enhances the efficiency
of the neutrino-heating mechanism. Therefore the 
multi-dimensional situation is {\em generically different} from 
the spherically symmetric case.

Nevertheless, the existence of convective overturn in the
neutrino-heating layer does not guarantee
explosions (\cite{janmue96,mezcal98:ndconv}).
For insufficient neutrino heating the threshold to an 
explosion will not be overcome. Since neutrinos play a crucial
role, an accurate description of the neutrino physics ---
transport and neutrino-matter interactions --- is indispensable
to obtain conclusive results about the viability of the 
neutrino-driven mechanism. All previously published multi-dimensional 
explosion models, however, have employed some crude approximations
or simplifications in the treatment of neutrinos.

\section{A New Generation of Multi-Dimensional Supernova Simulations}
\label{sec:results}

In order to take a next step of improvement in supernova
modelling, we have coupled a new Boltzmann solver for
the neutrino transport to the PROMETHEUS hydrodynamics code.
The combined program is called 
MuDBaTH ({\bf Mu}lti-\-{\bf D}i\-men\-sional {\bf B}oltzm{\bf a}nn
{\bf T}ransport and {\bf H}ydrodynamics) and allows for
spherically symmetric as well as multi-dimensional
simulations (\cite{ramjan02}).
Below we present some results of 1D and our first 2D supernova
simulations with this new code.

\subsection{Technical Aspects and Input Physics}
\label{sec:code}

For the integration of the equations of hydrodynamics we employ the
Newtonian finite-volume code PROMETHEUS (\cite{frymue89}), which was
supplemented by additional
problem specific features (\cite{kei97}).
PRO\-ME\-THE\-US is a direct Eulerian, time-explicit implementation of the
Piecewise Pa\-ra\-bol\-ic Method (PPM) of \cite{colwoo84}.
As a second-order Godunov scheme employing a Riemann solver it
is particularly well suited for following discontinuities in the
fluid flow like shocks or boundaries between layers of different
chemical composition. A notable advantage in the present context
is its capability of tackling multi-dimensional problems with high
computational efficiency and numerical accuracy. Our code
makes use of the ``Consistent Multifluid
Advection (CMA)'' method (\cite{plemue99}) for ensuring
accurate advection of different chemical components in the fluid,
and switches from the original PPM method to the more diffusive HLLE
solver of \cite{ein88} in the vicinity of strong shocks to avoid
spurious oscillations (the so-called ``odd-even decoupling''
phenomenon) when such shocks are aligned with one of the coordinate
lines in multidimensional simulations (\cite{qui94,kif00,plemue01}).

The Boltzmann solver scheme
is described in much detail elsewhere \hbox{(\cite{ramjan02})}. The
integro-differential character of the Boltzmann equation is
tamed by applying a variable Eddington factor closure to the
neutrino energy and momentum equations (and the simultaneously
integrated first and second order moment equations for neutrino
number). For this purpose the variable Eddington factor
is determined from the solution of the Boltzmann equation, and
the system of Boltzmann equation and its moment equations is
iterated until convergence is achieved. Employing this scheme in
multi-dimensional simulations in spherical coordinates, we solve the
radius- (and energy-) dependent
moment equations on the different angular bins of the numerical grid but
calculate the variable Eddington factor only once on an angularly
averaged stellar background. 
This approximation is good only for
situations without significant global deformations. Since
the iteration of the Boltzmann equation has to be done only
once per time step, appreciable
amounts of computer time can be saved (\cite{ramjan02}).
We point out here that it turned out to be
necessary to go an important step beyond this simple ``ray-by-ray''
approach. Physical constraints, namely the conservation of lepton
number and entropy within adiabatically moving fluid elements,
and numerical requirements, i.e., the stability of regions
which should not develop convection according to a mechanical stability
analysis, make it necessary to take into account the coupling of
neighbouring rays at least by lateral advection terms and neutrino
pressure gradients (Buras et al., in preparation).

General relativistic effects are treated only approximately in
our code (\cite{ramjan02}). The current version contains a modification
of the gravitational potential by including correction terms due to
pressure and energy of the stellar medium and neutrinos, which
are deduced from a comparison of the Newtonian and relativistic
equations of motion. The neutrino transport contains gravitational   
redshift and time dilation, but ignores the distinction between
coordinate radius and proper radius. This simplification is
necessary for coupling
the transport code to our basically Newtonian hydrodynamics.
Although a fully relativistic treatment would be preferable, tests
showed that these approximations seem to work satisfactorily well 
(Liebend\"orfer et al., in preparation), at least as
long as there are only moderate ($\sim$10--20\%) deviations of
the metric coefficients from unity and the infall velocities do
not reach more than 10--20\% of the speed of light in decisive
phases of the evolution.

As for the neutrino-matter interactions, we discriminate between
two different sets of input physics. On the one hand we have
calculated models with conventional (``standard'') neutrino
opacities, i.e.,
a description of the neutrino interactions which follows closely
the one used by Bruenn and Mezzacappa and collaborators
(\cite{bru85,mezbru93:coll,mezbru93:nes}). It assumes
nucleons to be uncorrelated, infinitely massive scattering
targets for neutrinos. In these reference runs we have
usually also added
neutrino pair creation and annihilation by nucleon-nucleon
bremsstrahlung (\cite{hanraf98}). Details of our
implementation of these neutrino processes can be found 
elsewhere (\cite{ramjan02}).

A second set of models was computed with
an improved description of neut\-rino-\-matter interactions.
Besides including nucleon thermal motions and recoil, which
means a detailed treatment of the reaction kinematics
and allows for an accurate evaluation of nucleon phase-space
blocking effects, we take into account nucleon-nucleon
correlations (following \cite{bursaw98,bursaw99}),
the reduction of the nucleon effective mass, and the
possible quenching of the axial-vector coupling
in nuclear matter (\cite{carpra02}). In addition, we have
implemented weak-magnetism corrections as described by
\cite{hor02}. The sample of neutrino processes was
enlarged by also including scatterings of muon and tau
neutrinos and antineutrinos off electron neutrinos and
antineutrinos and pair annihilation reactions between
neutrinos of different flavors
[i.e., $\nu_{\mu,\tau} + \bar\nu_{\mu,\tau}
\longleftrightarrow \nu_e + \bar\nu_e$; \cite{burjan02:nunu}].

Our current supernova models are calculated with the nuclear
equation of state of \cite{latswe91} using
an incompressibility modulus of $K = 180\,$MeV (other values
of the incompressibility of bulk nuclear matter in the 
Lattimer \& Swesty EoS cause only minor differences, see
\cite{thoetal02}).
For the density regime below $6\times 10^7\,$g/cm$^3$
we switch to an equation of state that considers
electrons, positrons and photons, and nucleons and nuclei with
an approximative treatment for composition changes due to nuclear
burning and shifts of nuclear statistical equilibrium (\cite{ramjan02}).

\begin{table}[htb]
\caption{Input physics for our set of computed models. 
See Sect.~\ref{sec:code} for details.}
\setlength\tabcolsep{6pt} 
\begin{tabular}{lllllll}
\sphline
\it Model & \it Dim. & \it Gravity & $\nu$ \it Reactions & \it Transport & \it Wedge$^a$ & \it Zones\cr
\sphline
s15Nso\_1d.a        & 1D & Newtonian  & standard       & Case A  &                   &    \cr
s15Nso\_1d.b        & 1D & Newtonian  & standard       & Case B  &                   &    \cr
s15Nso\_2d.a        & 2D & Newtonian  & standard       & Case A  & $\pm 27^{\rm o}$  & 20 \cr
s15Gso\_1d.b        & 1D & approx. GR & standard       & Case B  &                   &    \cr
s15Gso\_1d.b$^\ast$ & 1D & approx. GR & standard$^b$   & Case B  &              &    \cr
s15Gio\_1d.a        & 1D & approx. GR & improved$^c$   & Case A  &                   &    \cr
s15Gio\_2d.a        & 2D & approx. GR & improved       & Case A  & $\pm 43.2^{\rm o}$& 32 \cr
s15Gio\_1d.b        & 1D & approx. GR & improved       & Case B  &                   &    \cr
s15Gio\_2d.b        & 2D & approx. GR & improved       & Case B  & $\pm 43.2^{\rm o}$& 32 \cr
\sphline
\end{tabular}
\begin{tablenotes}
$^a$ Angular wedge of the spherical coordinate grid around the equatorial plane.\\
$^b$ Calculation without neutrino-pair creation by nucleon-nucleon
bremsstrahlung.\\
$^c$ Calculation without the neutrino-antineutrino processes of
\cite{burjan02:nunu}.
\end{tablenotes}
\label{table:1}
\end{table}

\begin{figure}[htb!]
\begin{minipage}[t]{0.48\textwidth}
\begin{center}
\includegraphics[width=1.0\textwidth]{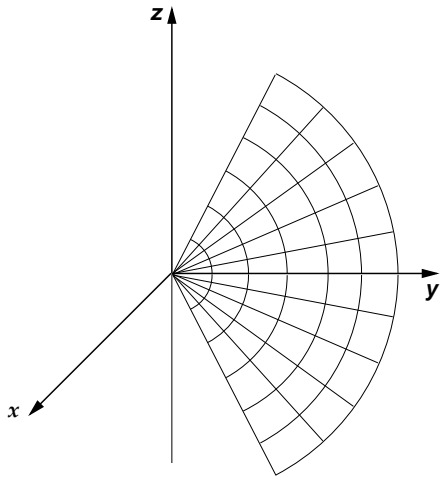}
\end{center}
\end{minipage}
\hspace{\fill}
\begin{minipage}[htp!]{0.45\textwidth}
\vspace{-4cm}
\caption[]{Computational grid for the 2D simulations.
         The wedge is placed around the equatorial plane
         of the spherical coordinate grid. Azimuthal symmetry
         around the $z$-axis is assumed, and periodic 
         conditions are used at the lateral grid boundaries.}
\label{fig:grid}
\end{minipage}
\end{figure}

\begin{figure}[htb!]
\begin{center}
\includegraphics[width=0.95\textwidth,clip=]{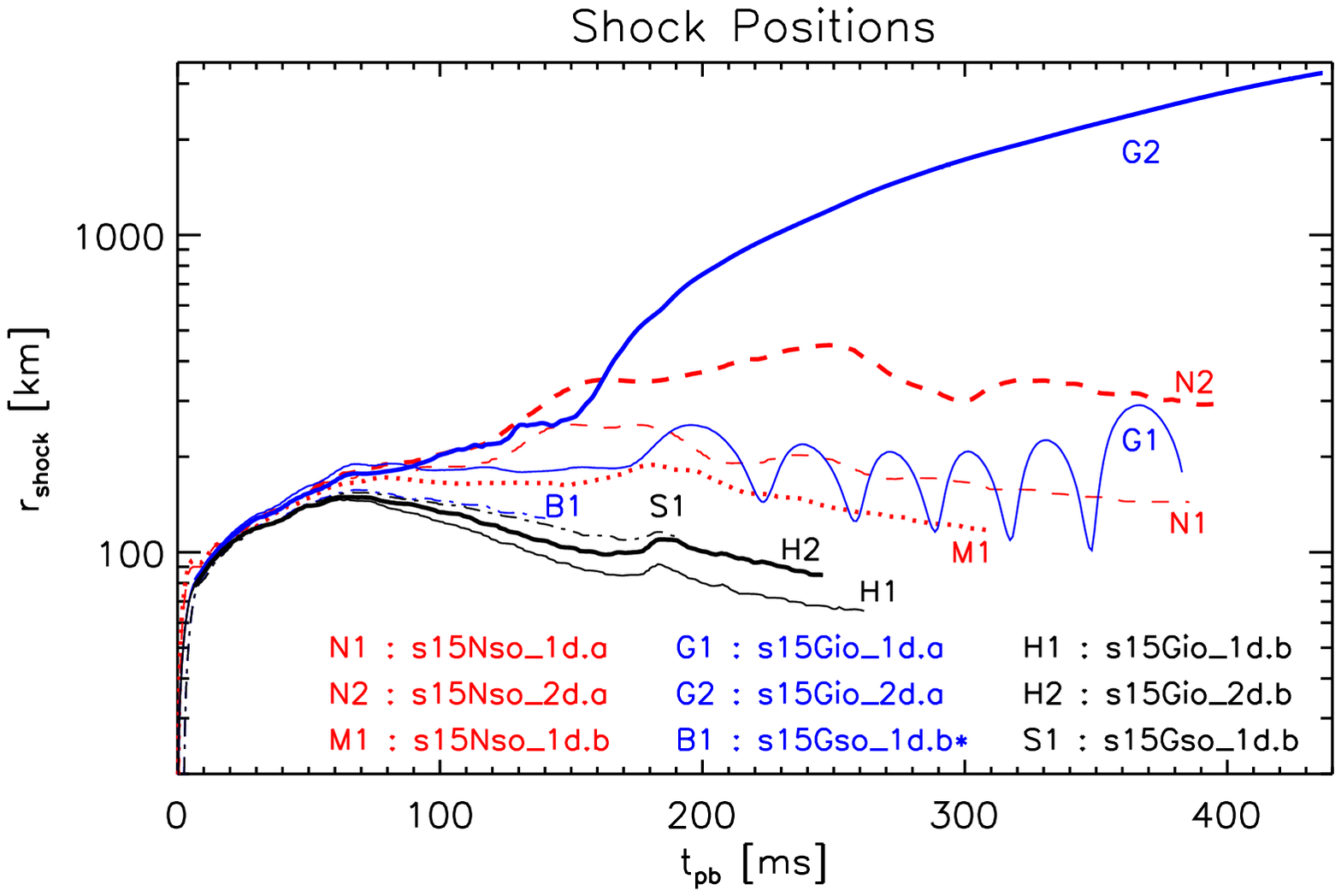}
\end{center}
\vspace{-3mm}
\begin{center}
\includegraphics[width=0.95\textwidth,clip=]{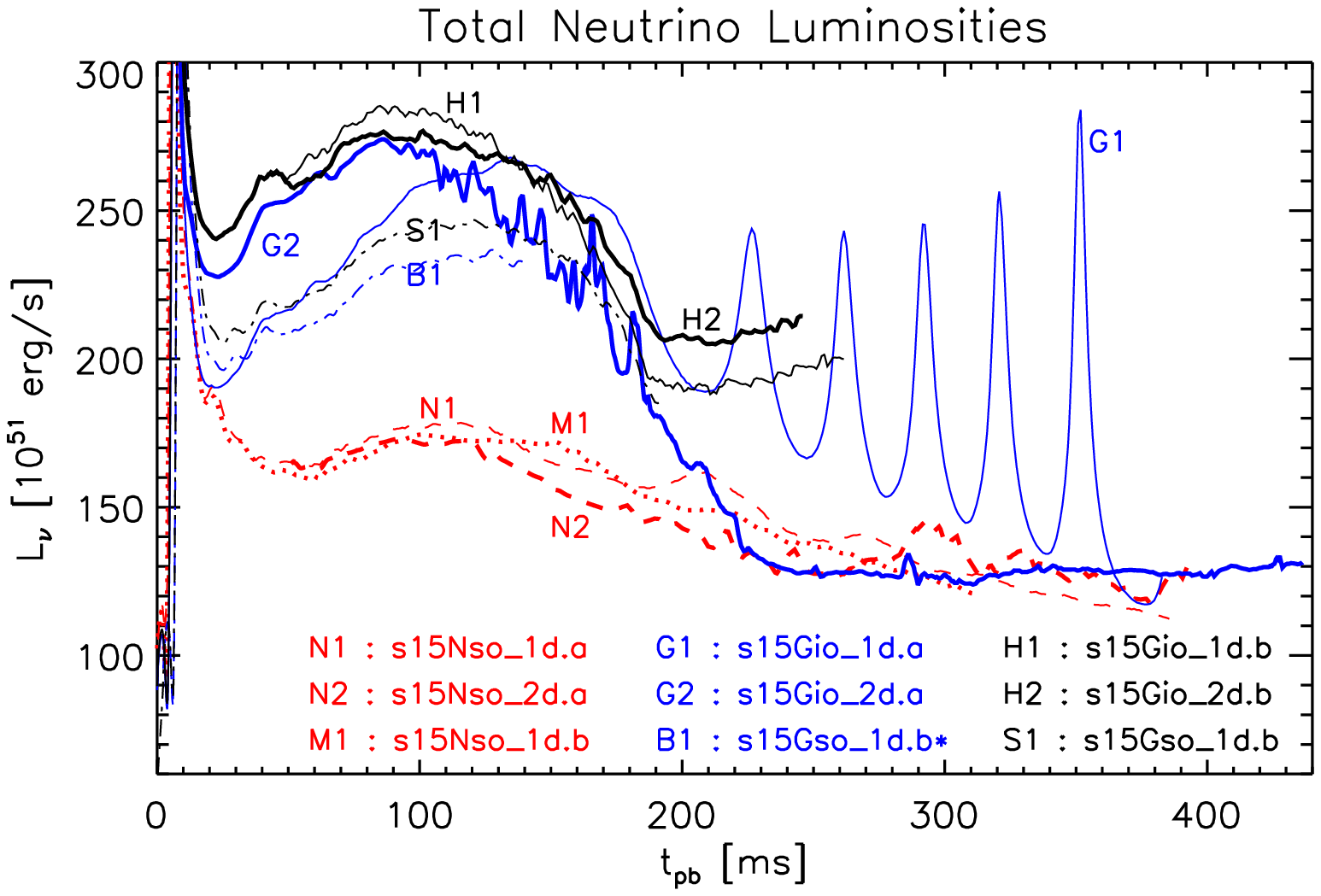}
\end{center}
\vspace{-3mm}
\caption[]{Shock trajectories (top) and total neutrino luminosities
(sum of luminosities of $\nu_e$, $\bar\nu_e$, and muon and
tau neutrinos and antineutrinos; bottom) for all models. 
Bold lines correspond to two-dimensional simulations.
Model~H2 (s15Gio\_2d.b) is the most complete 2D simulation. It
was computed with the improved set of neutrino opacities, relativistic
effects, and all velocity-dependent terms retained in the neutrino
transport equations.}  
\label{fig:shockradii}
\end{figure}

\begin{figure}[htb!]
\begin{center}
\includegraphics[width=0.915\textwidth,clip=]{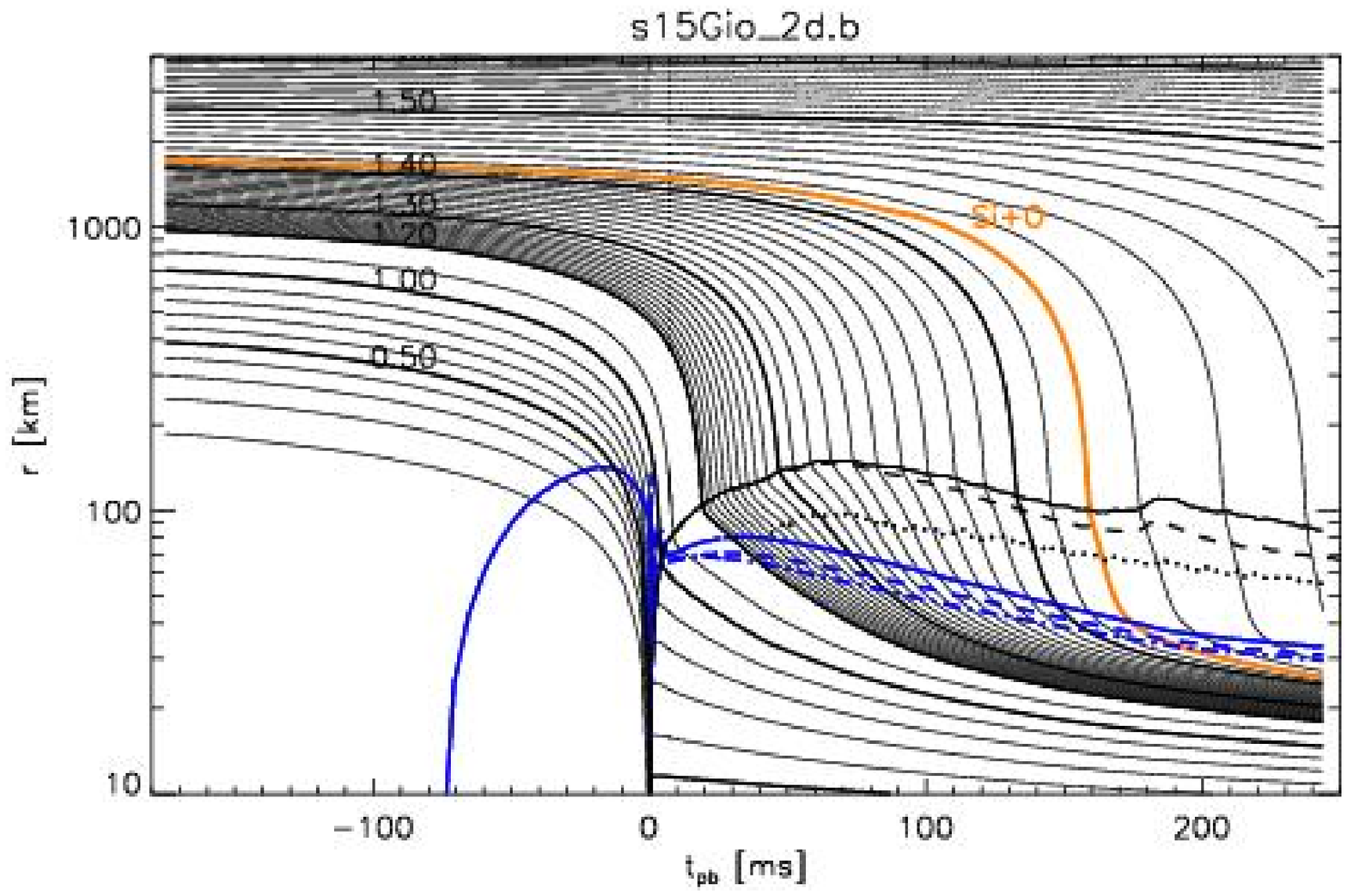}
\end{center}
\vspace{-3mm}
\begin{center}
\includegraphics[width=0.915\textwidth,clip=]{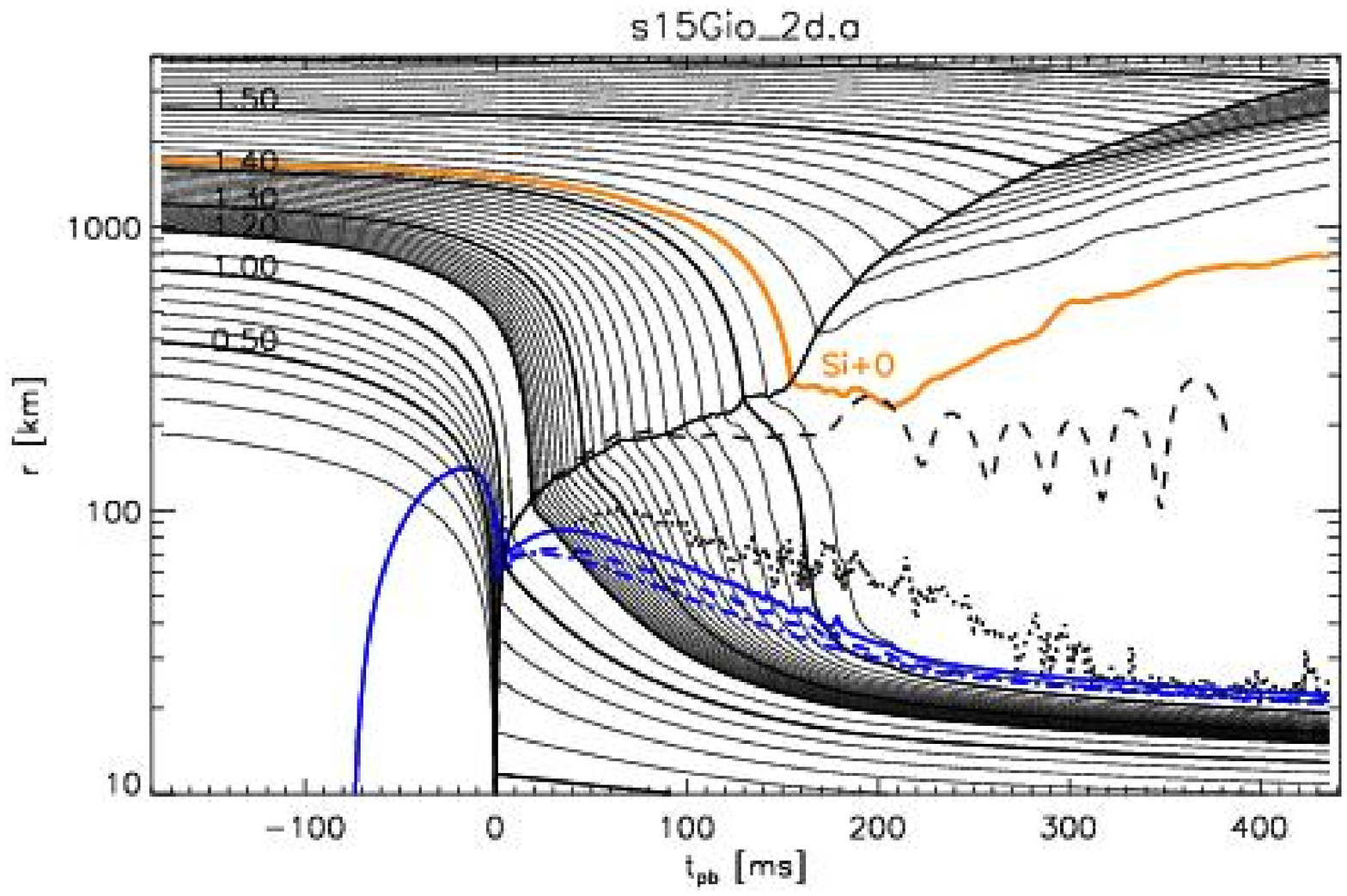}
\end{center}
\caption[]{Trajectories of mass shells (time being
normalized to bounce) for the
non-exploding (top) and the exploding 2D model.
In the latter case one can see the
shock starting a rapid expansion at about 150$\,$ms after
bounce. The dashed lines indicate the shock positions in
the corresponding 1D simulations, where no explosions were
obtained. The angle-averaged gain radius is given by the
dotted line, and the neutrinospheres of $\nu_e$, $\bar\nu_e$
and heavy-lepton neutrinos (lower solid, dashed, and
dash-dotted lines, respectively) are also marked.}
\label{fig:massshells}
\end{figure}

\subsection{Models and Results}
\label{sec:models}

We have performed a number of core-collapse simulations
in spherical symmetry and then followed the post-bounce
evolution in one and two dimensions. All described
calculations were started from a 15$\,$M$_{\odot}$ progenitor
star, Model s15s7b2, provided to us by S.~Woosley.
Adopting the naming, we label our models by s15N for
Newtonian runs and s15G for runs with approximate treatment
of general relativity, followed by letters ``so'' when  
``standard neutrino opacities'' were used and by ``io''
in case of our state-of-the-art improvement of the
description of neutrino-matter interactions. The model names
have a suffix that discriminates between 1D (``\_1d'') and
2D simulations (``\_2d''). For the 2D models we used a spherical 
coordinate grid with 20 or 32 equidistant zones within an angular
wedge from $-27^{\mathrm{o}}$ to $+27^{\mathrm{o}}$ or
from $-43.2^{\mathrm{o}}$ to $+43.2^{\mathrm{o}}$, respectively,
around the equatorial plane and assumed periodic conditions
at the boundaries~(Fig.~\ref{fig:grid}).
            
We have varied yet another aspect in our simulations.
Some of the models were computed with a version of the transport
code where the velocity dependent (Doppler shift and aberration)
terms in the neutrino momentum equation (and the corresponding
terms in the Boltzmann equation for the antisymmetric average
of the specific intensity; see \cite{ramjan02}) were
omitted. These terms are formally of order $v/c$ and should
be small for low velocities. This simplification of the
neutrino transport, our so-called ``Case A'',
was used in the models which have names ending with
the letter ``a''. Models with neutrino transport including
all velocity dependent terms also in the neutrino momentum
equation (``Case B'') can be identified by the ending ``b''
of their names. Table~\ref{table:1} provides an overview of   
the computed models.

The shock trajectories and corresponding total neutrino luminosities 
of all models are displayed in Fig.~\ref{fig:shockradii}.
No explosions, neither with the
standard nor with the improved description of the neutrino
opacities, were obtained with the
most complete implementation of the transport equations, Case~B
(see lines labeled with M1, H1, H2, S1, B1, corresponding to
Models~s15Nso\_1d.b, s15Gio\_1d.b, s15Gio\_2d.b, s15Gso\_1d.b and
s15Gso\_1d.b$^\ast$, respectively). The shock trajectories of
this sample of models form a cluster that is clearly separated
from the models computed with the transport version of Case~A, which
generally show a larger shock radius and therefore more optimistic
conditions for explosions. 

The simplification of the neutrino transport in Case~A
has a remarkable consequence which is obvious from
a comparison of 
the 2D runs of Models s15Gio\_2d.a and s15Gio\_2d.b,
which both were performed with our approximation of 
relativistic effects and the state-of-the-art improvement
of neutrino-matter interactions (cf.~Sect.~\ref{sec:code}).
While Model s15Gio\_2d.b fails to explode, the stalled shock in
Model s15Gio\_2d.a is successfully revived by
neutrino heating because very strong convection can
develop in the gain region.
The time evolution of both models is displayed 
by the trajectories of mass shells in
Fig.~\ref{fig:massshells}. 

\begin{figure}[!htbp]
\begin{center}
\includegraphics[width=0.9\textwidth,clip=]{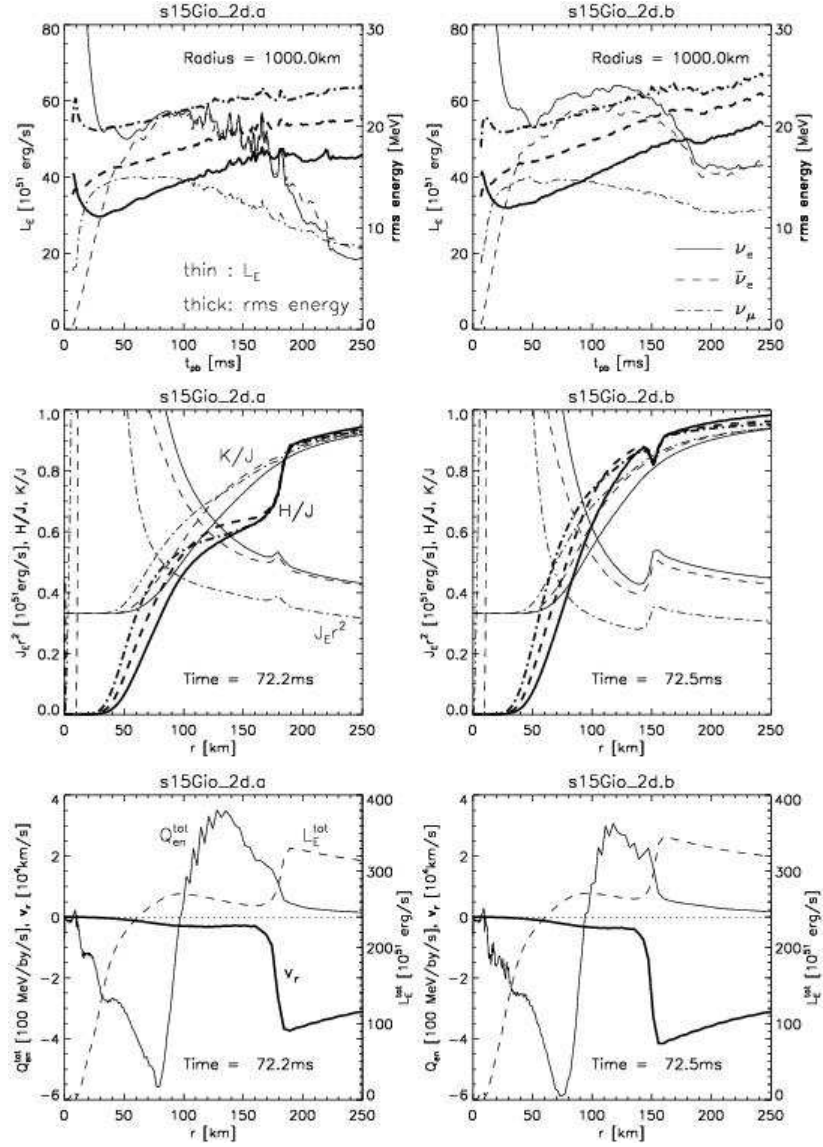}
\end{center}
\caption[]{Comparison of the exploding 2D model (Model~s15Gio\_2d.a; 
left) and the non-exploding 2D model (Model~s15Gio\_2d.b; right).
The top panels show the (comoving-frame) neutrino luminosities
and rms energies as functions of time after bounce.
The middle panels display the neutrino energy moment $J$,
multiplied by $r^2$ and the flux factor $H/J$ and Eddington
factor $K/J$ for both models at post-bounce time $t = 72.5\,$ms,
which corresponds roughly to the moment of maximum shock expansion
in Model~s15Gio\_2d.b (cf.\ Fig.~\ref{fig:massshells}).
The lower panels give the total neutrino
luminosities and neutrino heating rates in the comoving frame of
the stellar fluid at the same time.}
\label{fig:compare}
\end{figure}

The reason for this dramatic difference can be understood
in the following way.
Some of the velocity dependent terms (those in which
derivatives with respect to the neutrino energy do not
show up) in the neutrino momentum equation
have a simple formal interpretation: In regions
with mass infall ({\em negative} velocity) they effectively
act like a {\em reduction}
of the neutrino-medium interaction on the right hand
side of this equation. The corresponding changes can be 10\% 
or more for neutrino energies in the peak of the spectrum,
depending also on time, radius, and the size of the
postshock velocities. As a consequence of the omission of 
these terms in Model~s15Gio\_2d.a, neutrinos are impeded
in their streaming. Therefore the comoving-frame neutrino
(energy) density is increased (see Fig.~\ref{fig:compare}).
This is associated with slightly lower neutrino losses
in the cooling layer around the neutrinosphere and a 
significantly enhanced neutrino heating between 
gain radius and shock.

\begin{figure}[!htbp]
\begin{center}
\includegraphics[width=0.9\textwidth,clip=]{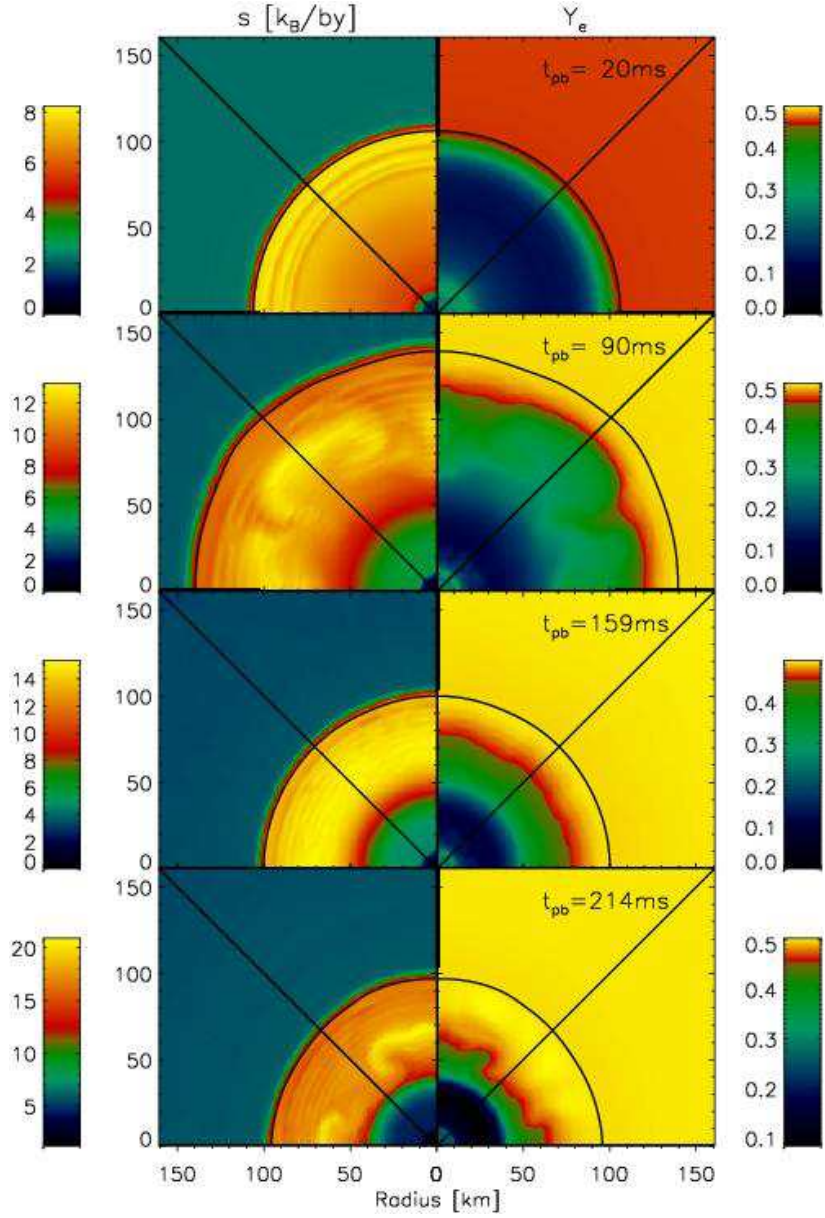}
\end{center}
\caption[]{Convection in the neutrino-heating region for
the non-exploding 2D model (Model~s15Gio\_2d.b) at the
post-bounce times indicated in the plots. The figures show
the entropy distribution (left) and the
electron fraction (proton-to-baryon ratio). A wedge of
$\pm 43.2^{\mathrm o}$ around the equatorial plane (marked
by the diagonal solid lines) of the spherical coordinate
grid was used for the computation.}
\label{fig:snapshots1}
\end{figure}

\begin{figure}[!htbp]
\begin{center}
\includegraphics[width=0.9\textwidth,clip=]{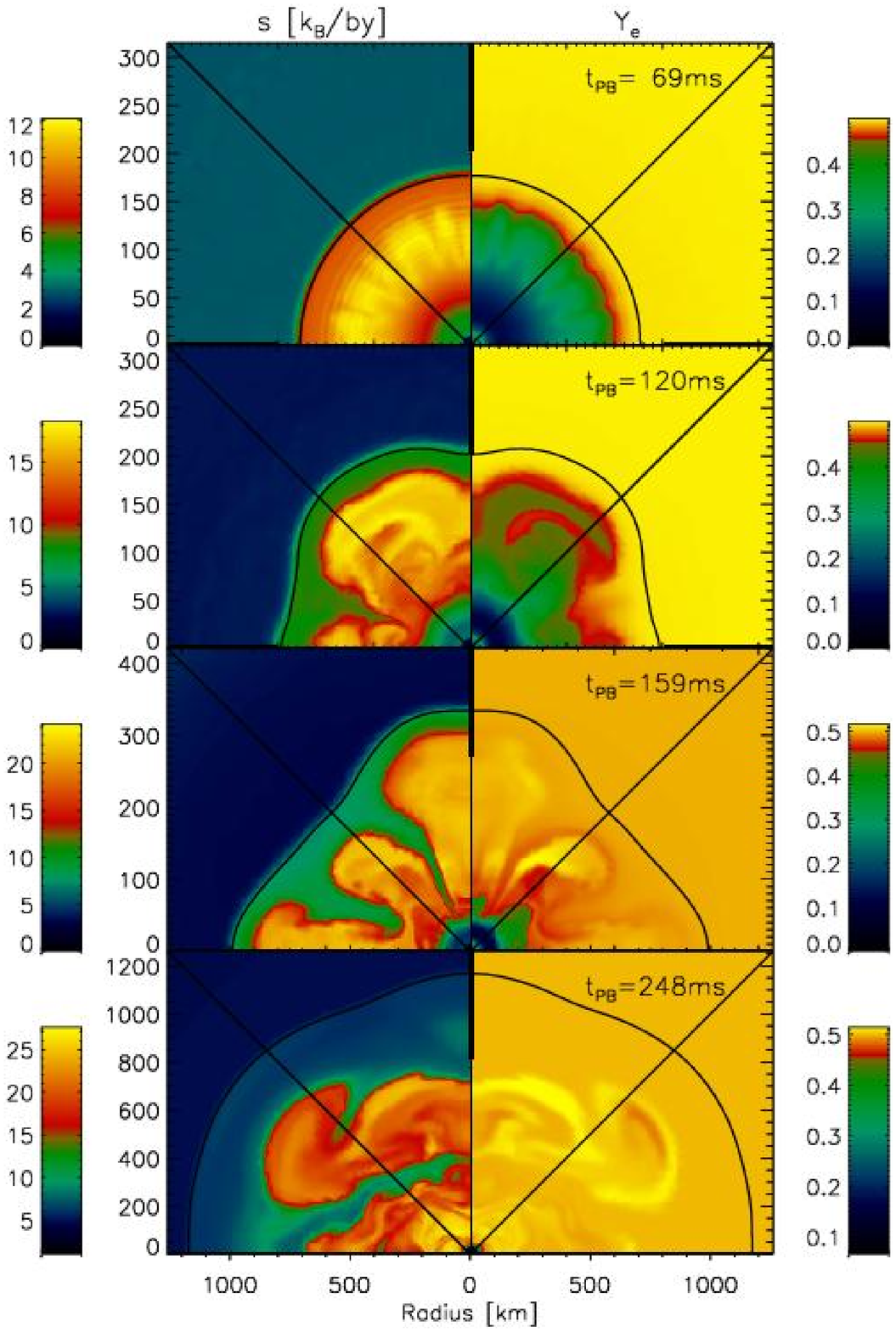}
\end{center}
\caption[]{Same as Fig.~\ref{fig:snapshots1}, but for the
exploding 2D model, Model~s15Gio\_2d.a. Note the different
radial scales of the four snapshots.}
\label{fig:snapshots2}
\end{figure}

Conversely, the inclusion of the velocity-dependent terms in the
neutrino momentum equation has negative consequences for the
shock propagation in Model s15Gio\_2d.b. 
Although the differences at early post-bounce times are moderate
(10--30\%, depending on the quantity) 
the accumulating effects during the first 80$\,$ms after bounce
clearly damp the shock expansion and finally lead to a dramatic 
shock recession after the initial phase of expansion.
Before this happens postshock convection has not become strong
enough to change the evolution. With the onset of contraction, 
the postshock velocities decrease (become more negative)
quickly, neutrino-heated
matter is rapidly advected down through the gain radius and loses
its energy by the reemission of neutrinos. The larger negative
values of the postshock velocity enhance the influence of the
velocity-dependent terms and reduce the heating efficiency even
more. Because of this disastrous feedback, the gain region
shrinks to a very narrow layer, a fact which suppresses 
the convective activity lateron. This is 
demonstrated by Figs.~\ref{fig:snapshots1} 
and \ref{fig:snapshots2}, which show that the convection
is weak in Model~s15Gio\_2d.b but very strong in 
Model~s15Gio\_2d.a. 

Due to a combination of unfavorable effects and a continuously
amplifying negative trend, Model~s15Gio\_2d.b remains below the 
explosion threshold while Model~s15Gio\_2d.a is just above 
that critical limit. In the vicinity of the threshold the
long-time evolution of the collapsing stellar core therefore 
depends very sensitively on ``smaller details'' of the neutrino
transport.

The two simulations also demonstrate that a
sufficiently large shock radius
for a sufficiently long time is crucial for the 
growth of convective overturn in the neutrino-heating layer.
Only when convection behind the shock becomes strong it can
be decisive for getting an explosion. This latter fact is known
from previous multi-dimensional 
simulations (\cite{janmue96,mezcal98:ndconv}) and is
again confirmed by a comparison of Model~s15Gio\_2d.a with
the corresponding one-dimensional simulation, Model~s15Gio\_1d.a.
It is interesting that the Newtonian 2D run, Model~s15Nso\_2d.a
(with standard opacities) does marginally fail, whereas
Model~s15Gio\_2d.a with relativistic corrections (and
state-of-the-art neutrino reactions) succeeds. The influence
of the neutrino opacities can be directly seen by
comparing Models~s15Gso\_1d.b and s15Gio\_1d.b. It is not
dramatic, but sufficient to justify the inclusion of the
improvements described in Sect.~\ref{sec:code}.
Neutrino-pair creation by bremsstrahlung makes a minor
difference during the considered phases of the evolution
(Model~s15Gso\_1d.b$^\ast$ vs.\ Model~s15Gso\_1d.b).

\begin{figure}[htb!]
\begin{center}
\includegraphics[width=1.0\textwidth,clip=]{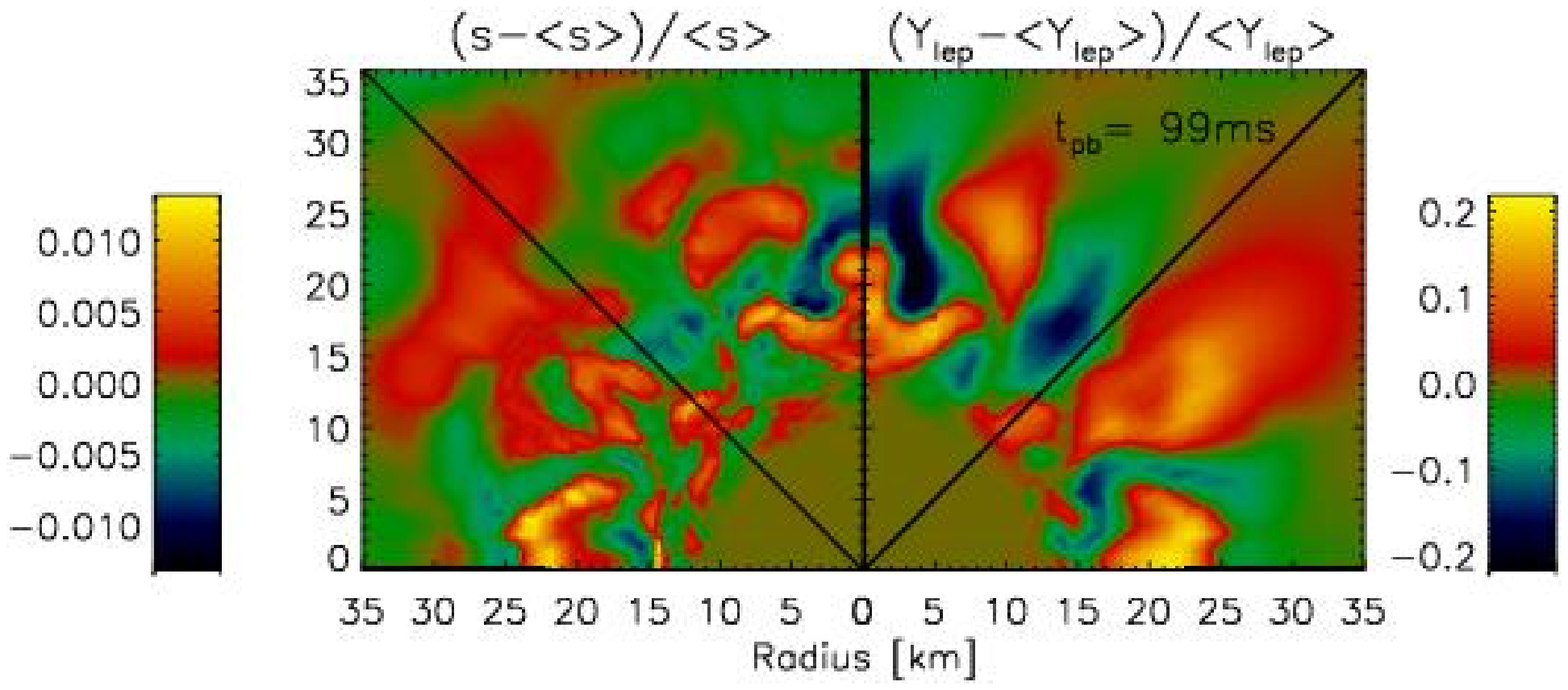}
\end{center}
\vspace{-3mm}
\begin{center}
\includegraphics[width=0.75\textwidth,clip=]{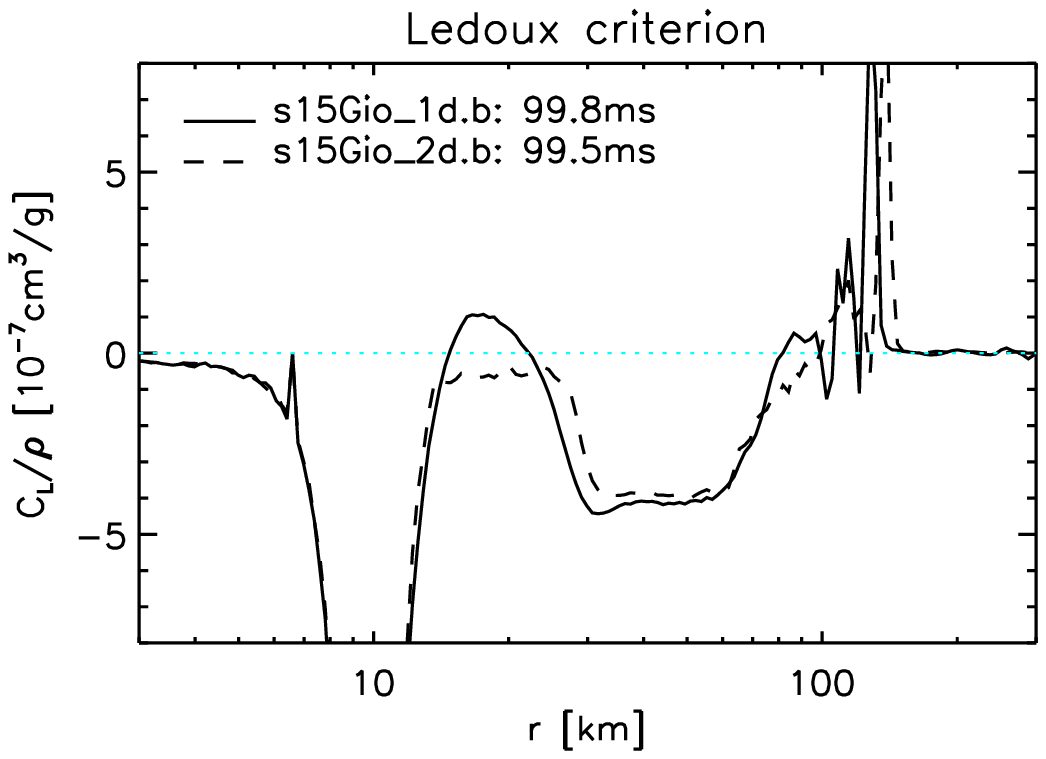}
\end{center}
\caption[]{Convection below the neutrinosphere at about 100$\,$ms
after core bounce in Model~s15Gio\_2d.b. The upper figure
shows the relative fluctuations of entropy (left) and 
lepton fraction (right), normalized to the mean values at
a given radius. The lower figure displays the Ledoux criterion
for the 1D Model~s15Gio\_1d.b (without convection; solid line) 
and the 2D Model~s15Gio\_2d.b (dashed line). Positive values indicate
convectively unstable conditions, once in an inner region 
(located at about 15--30$\,$km) below the neutrinosphere and another
time in the neutrino-heating layer behind the shock.}
\label{fig:nsconvection}
\end{figure}

We also found that Ledoux convection sets in {\em below} the 
neutrinosphere already shortly after bounce
(Fig.~\ref{fig:nsconvection}). It is 
persistent until the end of our simulations and slowly digs
farther into the star as already found in the simulations
of \cite{keietal96}. But the convective activity is
so deep inside the neutron star that its effects on the
$\nu_e$ and $\bar\nu_e$ luminosities and on the supernova
dynamics are insignificant.

\begin{figure}[htb!]
\begin{minipage}[t]{0.48\textwidth}
\begin{center}
\includegraphics[width=1.0\textwidth]{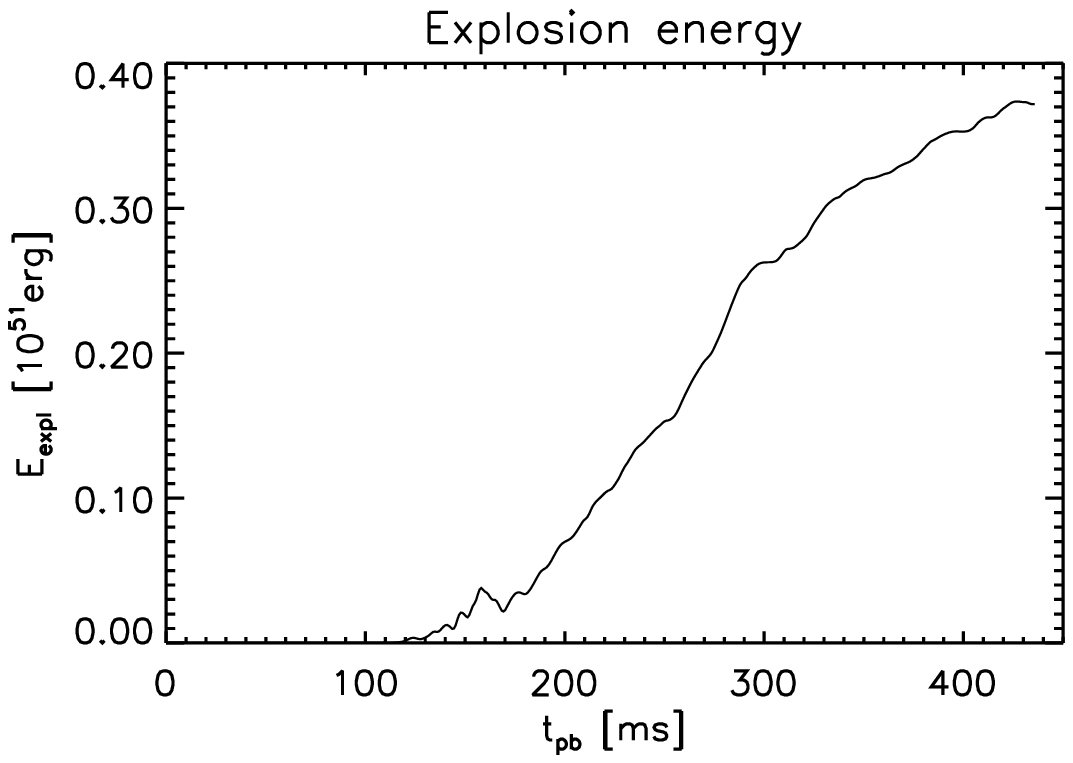}   
\end{center}
\vspace{-3mm}
\caption[]{Explosion energy as function of time for Model~s15Gio\_2d.a.}
\label{fig:explosionenergy}
\end{minipage}
\hspace{\fill}
\begin{minipage}[t]{0.48\textwidth}
\begin{center}
\includegraphics[width=1.0\textwidth]{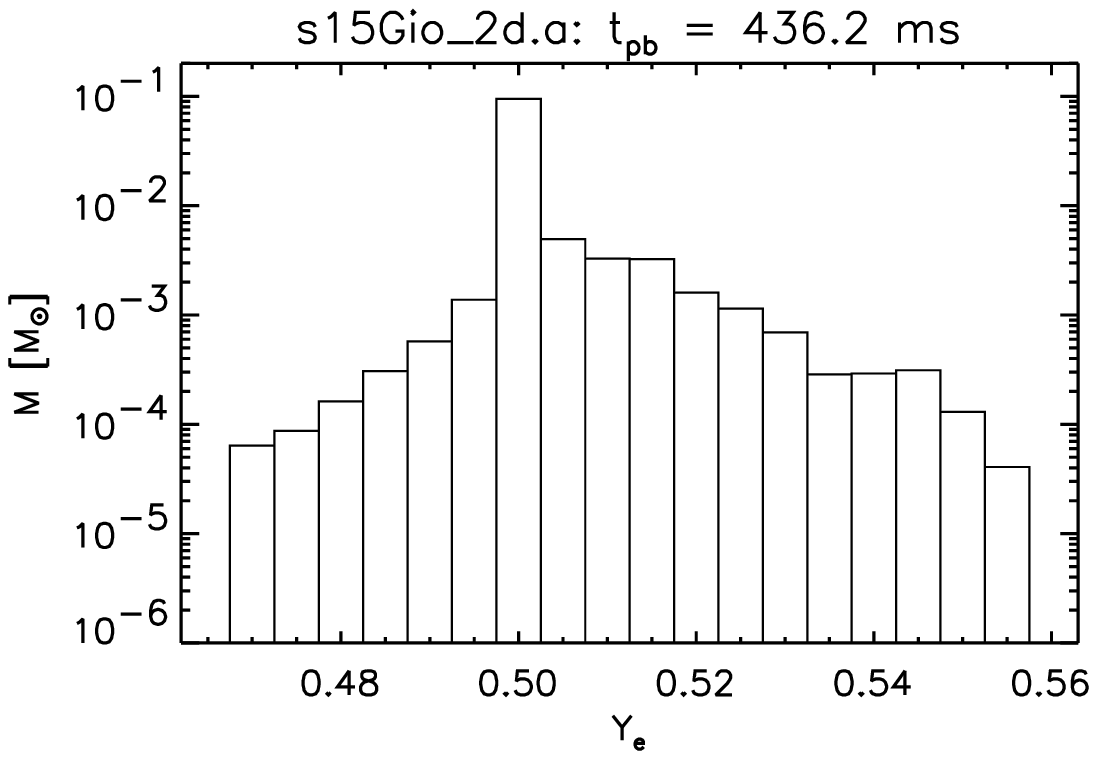}
\end{center}
\vspace{-3mm}
\caption[]{Ejecta mass vs.~$Y_e$ for Model s15Gio\_2d.a at 436$\,$ms
         after bounce. The total mass of $^{56}$Ni is about 
         0.1$\,$M$_{\odot}$.}
\label{fig:yemass}
\end{minipage}
\end{figure}

\section{Conclusions and Outlook}
\label{sec:conclusions}

We believe that our 2D models with a Boltzmann solver
for the neutrino transport and a state-of-the-art
description of neutrino-matter interactions have 
considerably reduced the
uncertainties associated with the treatment of the
neutrino physics in previous multi-dimensional 
simulations. With the most complete implementation of
the transport physics we could not obtain explosions.
This result suggests that the neutrino-driven
mechanism fails with the employed input physics,
at least in case of the considered 15$\,$M$_{\odot}$
star. We do not think that the 
approximate treatment of general relativistic effects
is likely to jeopardize this conclusion. A comparison
with fully relativistic one-dimensional calculations
(Liebend\"orfer et al., in preparation) 
is very encouraging. Because of the remarkable similarity 
of the shock trajectories of different progenitors 
in spherical symmetry (\cite{liemes02}), it is probable
that our negative conclusion is also valid for other
pre-collapse configurations with a similar structure.
Significant star-to-star variations of the 
progenitor properties with a non-monotonic dependence 
on the stellar mass (\cite{wooheg02}),
however, suggest that 
multi-dimensional core-collapse simulations of a
larger sample of stars are needed before one can 
make final, more generally valid statements.

Another concern may be the omission of lateral neutrino
fluxes in our multi-dimensional transport approximation.
We describe the neutrino losses from the hot spots
that appear at locations where the low-entropy downflows
plunge into the neutrinospheric layer, by just a radial
flux. In reality the neutrinos produced in this
semi-transparent matter would be radiated away more 
isotropically, a fact which could lead to a larger
reabsorption probability in the surrounding medium. The
corresponding increase of the heating efficiency might not
be very large, but could still be relevant because of the
close proximity of our models to an explosion.
The supernova problem is highly nonlinear 
and surprises may lurk behind every corner.

It would therefore be premature to conclude that the
neutrino-driven mechanism fails and that not even 
postshock convection can alter this unquestioned 
outcome of all current spherical models. Besides 
studying other progenitors with multi-dimensional 
simulations, one should also investigate the effects
of rotation during the neutrino-heating phase and
long-time post-bounce evolution of a supernova.
Previous simulations found weaker explosions in case
of rotating models (\cite{fryheg00}). But
these calculations were performed with a much simplified
treatment of the neutrinos and showed powerful, nearly
prompt explosions already in the absence of rotation.
It cannot be excluded that even a moderate amount of
rotation might have significant consequences when
added to nonrotating models
which explode only with a long time delay after core
bounce or even fail to do so.
Naturally, three-dimensional simulations are ultimately
desirable and could reveal generic differences compared
to 2D, in particular in combination with rotation. The
recent results of \hbox{(\cite{frywar02})}, which
suggest a large degree of similarity between 2D and 3D
results, are groundbreaking but again must be
interpreted with caution. The serious approximations
of the neutrino physics and the rapid development of
the explosion in these models do not allow too 
far-reaching conclusions.

Another major uncertainty of current supernova models
is associated with the incompletely
known physics in the nuclear and supranuclear medium.
Not only does the high-density EoS affect the core bounce,
shock formation, and prompt shock propagation, it also
determines the internal properties of the hot neutron star
and thus
its neutrino emission, both indirectly and directly.
Hadronic degrees of freedom in addition to nucleons in the
supranuclear matter, e.g., hyperons, pions or kaons,
can reduce the stiffness of the EoS at very high densities.
The possible effects on the supernova explosion are
essentially unexplored. A softening of the EoS can, for
example, lead to a faster contraction and heating
of the nascent neutron star. This in turn also determines
the density and temperature profile in the vicinity of the
neutrinosphere where most of the neutrino flux is produced
during the crucial phase of delayed shock revival.

Our successfully exploding 2D model, Model~s15Gio\_2d.a, 
at least demonstrates that simulations which include the
effects of postshock convection are rather close
to an explosion. Therefore modest changes of the neutrino 
emission and transport (in our case the omission of the
velocity-dependent terms in the neutrino-momentum equation)
seem to be already sufficient to push 
them beyond the critical
threshold. The properties of the explosion in this case are
very encouraging and may support one's belief in the basic 
viability 
of the delayed explosion mechanism. At 440$\,$ms after bounce 
the shock has arrived at a radius of more than 3000$\,$km and
is expanding with about 10000$\,$km/s (Fig.~\ref{fig:shockradii}).
The explosion of this model does not become very energetic
(Fig.~\ref{fig:explosionenergy}). The energy
is just $\sim 4\times 10^{50}\,$erg at that time and 
grows only very slowly. This may not be a serious problem
if one recalls the large spread of energies
of observed supernovae (Supernova~1999br, for example, 
is estimated to have an ejecta mass of $14\,$M$_{\odot}$
and an explosion energy of about $6\times 10^{50}\,$erg; \cite{ham02}).
Moreover, the explosion energy is a very steep function of the
physical conditions in models which are above the explosion
threshold (\cite{janmue95}). A significant increase of
the energy may therefore not require large changes.

Since the explosion starts rather late (at $\sim 150\,$ms post 
bounce), the proto-neutron star has accreted enough matter to have  
attained an initial baryonic mass of 1.4$\,$M$_{\odot}$. Therefore
our simulation does not exhibit the problem of previous
successful multi-dimensional calculations which produced 
neutron stars with masses on the lower side of plausible
values ($\sim 1.1\,$M$_{\odot}$).
Also another problem of published explosion models
(e.g., \cite{herben94,burhay95,janmue96,fry99})
has disappeared: The ejecta mass with $Y_e \la 0.47$
is less than $\sim 10^{-4}\,$M$_{\odot}$ in our
calculation (Fig.~\ref{fig:yemass}),
thus fulfilling a constraint pointed out by
\cite{hofwoo96} for supernovae
if they should not overproduce the $N=50$ (closed neutron
shell) nuclei, in particular $^{88}$Sr, $^{89}$Y and $^{90}$Zr,
relative to the Galactic abundances. Of course, final statements
about explosion energy, ejecta composition, and the neutron star
mass (which may change due to later fallback, especially when the 
explosion energy remains low) require to follow the evolution for
a longer time. 

\section*{Final Words}

There is still a long way to go until we will have arrived
at a ``standard model'' for massive star explosions.
Current numerical simulations are not advanced enough to
convincingly and self-consistently demonstrate the viability 
of any of the explosion scenarios described in 
Sect.~\ref{sec:theory}. 
Despite of respectable progress in treating the neutrino 
transport, general relativity, and multi-dimensional aspects
in the most detailed numerical calculations, 
the models are far from
being complete. Neglecting potentially important physics like,
e.g., magnetic fields, one cannot be sure that one is searching
for the actual cause of massive star explosions at the right place. 
Conversely, the existing simulations also do not bring us in 
a position to reject the idea that neutrinos deliver the
required energy in the ejecta.

With the lack of direct observational information and being
confronted with the fact that our knowledge about the physics 
in collapsing stellar cores is incomplete, there is no 
alternative to systematically explore the space of possibilities
by hydrodynamical simulations. Since ``the output of complex 
numerical models is as inscrutable as nature herself''
\hbox{(\cite{ema91})}, it seems advisable to first develop reliable 
and accurate supernova models with a minimum of questionable
assumptions and then to move on step by step to expanding them 
towards larger complexity and more completeness.

For the time being,
one's favorate choice of the supernova mechanism is more a
matter of taste and belief than based on undisputable facts and logical
reasoning. The ``ifs'' and ``buts'' in current models are still too many.
One should be cautious not to put too much money
on any single possibility right now.

\begin{acknowledgments}
We are indebted to K.~Takahashi for
providing us routines to calculate the improved neutrino-nucleon
interactions, and to C.~Horowitz for correction formulae for the
weak magnetism.
We also thank M.~Liebend\"orfer for making output data of his 
simulations available to us for comparisons.
The Institute for Nuclear Theory at the University
of Washington is acknowledged for its hospitality and the
Department of Energy for support during a visit of the Summer Program
on Neutron Stars, during which most of the work leading to 
Fig.~\ref{fig:monomode} was done. HTJ, RB and MR
are grateful for support by the Sonderforschungsbereich
375 on ``Astroparticle Physics'' of the Deutsche Forschungsgemeinschaft.
TP was supported in part by the US Department of Energy
under Grant No. B341495 to the Center of Astrophysical Thermonuclear
Flashes at the University of Chicago, and in part by the grant
2.P03D.014.19 from the Polish Committee for Scientific Research.
He performed his simulations on the CRAY SV1-1A at the 
Interdisciplinary Centre for Computational Modelling in Warsaw.
The 2D simulations with Boltzmann neutrino transport were only 
possible because a node of the   
new IBM ``Regatta'' supercomputer was dedicated to this project
by the Rechenzentrum Garching. Computations were also
done on the NEC SX-5/3C
of the Rechenzentrum Garching, and on the CRAY T90 and CRAY
SV1ex of the John von Neumann Institute for Computing (NIC) in J\"ulich.
\end{acknowledgments}

\end{document}